\documentclass[aps,prd,amssymb,final,nobibnotes,nofootinbib,twocolumn,superscriptaddress]{revtex4-2}
\usepackage{graphicx}
\usepackage{hyperref}
\usepackage{amsmath}
\usepackage{color}
\usepackage{xcolor}
\usepackage{numprint}
\usepackage{float}
\usepackage{ulem}
\usepackage{physics}
\usepackage{mathtools}
\usepackage{dsfont}
\usepackage{subfigure}
\usepackage{upgreek}
\usepackage{multirow}
\usepackage{mathrsfs}

\begin{document}
\title{Semiclassical bremsstrahlung from a charge radially falling into a Schwarzschild black hole}

\author{Jo\~ao P. B. Brito}
\email{joao.brito@icen.ufpa.br} 
\affiliation{Programa de P\'os-Gradua\c{c}\~{a}o em F\'{\i}sica, Universidade 
		Federal do Par\'a, 66075-110, Bel\'em, Par\'a, Brazil}

\author{Rafael P. Bernar}
\email{rbernar@ufpa.br}
\affiliation{Programa de P\'os-Gradua\c{c}\~{a}o em F\'{\i}sica, Universidade 
		Federal do Par\'a, 66075-110, Bel\'em, Par\'a, Brazil}
		
\author{Atsushi Higuchi}
\email{atsushi.higuchi@york.ac.uk}
\affiliation{Department of Mathematics,  University of York, Heslington, York YO10 5DD, United Kingdom}

\author{Lu\'is C. B. Crispino}
\email{crispino@ufpa.br}
\affiliation{Programa de P\'os-Gradua\c{c}\~{a}o em F\'{\i}sica, Universidade Federal do Par\'a, 66075-110, Bel\'em, Par\'a, Brazil}%

\date{\today}
\begin{abstract}
A semiclassical investigation of the electromagnetic radiation emitted by a charged particle in a radially freely falling motion in Schwarzschild spacetime is carried out. We use quantum field theory at tree level to obtain the one-particle-emission amplitudes. We analyze and compare the energy spectrum and total energy released, which are calculated from these amplitudes, for particles with varying initial positions and for particles originating from infinity with varying kinetic energy. We also compare the results with those due to a falling charged ``string'' extended in the radial direction.

\end{abstract}

\maketitle

\section{Introduction}

The radio and gravitational wave astronomy has ushered in a new era in black hole (BH) physics~\cite{EHT_sombra,EHT_sombra_SgrA,ligo1_2016,ligo2_2016}, giving complementary experimental data to best test general relativity and alternative theories of gravity in a strong field regime~\cite{berti_2015,psaltis_2019,koyama_2016}, which is the regime where we are more likely to find deviations, if there are any, from the predictions of these theories. Moreover, the study of fundamental fields associated with dynamical processes near BHs, e.g., the radiation emitted by spiraling matter, plays a crucial role in high-energy astrophysics~\cite{Hoyle1963,Salpeter1964,Lynden-Bell1969,Kormendy2013}.
For example, the energetic events near the center of a Seyfert galaxy are widely believed to be due to its central supermassive BH intensely interacting with surrounding material~\cite{weedman_1977,hills_1975}. In particular, the radiation emitted to infinity by dynamical processes carries ``fingerprints'' of the BH and its vicinity~\cite{vishveshwara_1970,falcke_2000,wilkins_2021}.

For a full description of physics near BHs, the quantum nature of gravity must be taken into account~\cite{Turyshev2008,hawking_1996}. General relativity predicts the development of singularities in which the concepts of spacetime and matter break down, signaling the need for new physics at the Planck scale ($\sim 10^{-33}$ cm) where quantum gravity is expected to take over.
Although finding the full quantum theory of gravity describing nature remains an open problem in theoretical physics~\cite{kiefer_2005}, important results have been achieved with quantum field theory (QFT) in curved spacetime~\cite{birrell_1982,parker_2009}. 

Quantum field theory in curved spacetime emerged from an investigation of particle creation in expanding universes~\cite{parker_1969}. This theory deals with quantum fields in fixed background spacetimes and is a generalization of QFT in (flat) Minkowski spacetime.
Quantum field theory in curved spacetime gained impetus from the remarkable discovery using this theory that BHs radiate as black bodies (Hawking radiation), raising the possibility of their dramatic disappearance through this thermal radiation~\cite{hawking_1974,hawking_1975}. Soon after Hawking's discovery, Unruh published another remarkable result examining aspects of BH evaporation. His result reveals the observer-dependent nature of the particle content in field theory (Unruh effect)~\cite{fulling_1973,davies_1975,unruh_1976}. 
The semiclassical approach of QFT in fixed background spacetime, although it is only an effective theory, reveals aspects that a complete theory of quantum gravity must exhibit. The semiclassical results thus play a key role in any approach to quantum gravity~\cite{kiefer_2007}.

Black holes are believed to be surrounded by spiraling matter that forms accretion disks. The accretion of matter results in the release of gravitational potential energy, which is the main source of power in the center of galaxies~\cite{yuan_2014}. When matter falls into BHs, it emits radiation in various channels. This process was investigated in the $1970s.$ Using the formalism given by Regge and Wheeler~\cite{regge_1957} and by Mathews~\cite{mathews_1962}, Zerilli computed the gravitational radiation of a particle falling into a Schwarzschild BH~\cite{zerilli_1970}. Motivation to study radiative processes near BHs increased when Weber reported (now discredited) evidence for discovery of gravitational radiation~\cite{weber_1969} (see, e.g., Refs.~\cite{davis_1971,misner_1972,misner_et_al_1972,davis_1972} and the references therein). Further analyses of the gravitational radiation using classical field theory can be found in Refs.~\cite{davis_et_al_1972,ruffini_1973,ferrari_1981,lousto_1997,martel_2002,cardoso_2002}. For such analyses in Kerr BH spacetime, see Refs.~\cite{cardoso_2003,cardoso2_2003}. 
On the other hand the study of the electromagnetic radiation emitted in the vicinity of BHs may be used to test, e.g., the Kerr BH hypothesis~\cite{bambi_2017}. As for the electromagnetic radiation emitted by a charged particle in radial free fall, Ruffini \textit{et al.} computed the amount of energy and the spectral distribution~\cite{ruffini2_et_al_1972,ruffini_1972} (see also Ref.~\cite{tiomno_1972}). More recently, Cardoso \textit{et al.} have investigated the electromagnetic radiation emitted by an ultrarelativistic infalling charged particle~\cite{cardoso_eletr_2003}. Folacci and Ould El Hadj studied the electromagnetic radiation generated using the complex angular momentum description~\cite{folacci_2020}.

In this paper, we investigate the radiation emission phenomena considering QFT instead of classical field theory, i.e., by using QFT in curved spacetime at tree level, in the vicinity of
a nonrotating BH. In this approach,
the classical charge is coupled to the quantum field, giving rise to a nonvanishing one-particle-emission probability.
The quantization of the electromagnetic field in a curved background was performed,
e.g., in Refs.~\cite{higuchi_1987,crispino_1998,crispino_2001,castineiras_2005}.  
The scalar radiation emitted by a radially infalling source was investigated using QFT by Oliveira and some of the present authors~\cite{oliveira_2018}.
Although QFT at tree level yields the same results as classical field theory, it will give a different perspective and serve as a starting point for finding quantum corrections.
Here, we investigate the electromagnetic radiation emitted by charged particles freely falling radially into a Schwarzschild BH from some initial radial position
from rest. We also consider nonzero initial velocity for the case where the charged particle
falls from infinity. We use the test-particle approximation,
which is valid if the mass of the charged particle is much
smaller than the BH mass.
It is interesting that there is good agreement between this approximation and the numerical computation in the fully nonlinear regime of general relativity, e.g., for BH collisions~\cite{anninos_1993,anninos_1995,sperhake_2011}.

The rest of this paper is organized as follows.
In Sec.~\ref{field_quantization}, we review some general features of the electromagnetic field quantization in Schwarzschild spacetime. In Sec.~\ref{radiation_emission}, we calculate the one-particle-emission amplitude and study the radiation emitted by the infalling charged particle. In Sec.~\ref{low_frequency}, we 
find the zero-frequency limit of some electromagnetic energy spectra
analytically and compare them with the corresponding numerical results.
In Sec.~\ref{sec_results}, we plot some selected numerical results and give our final remarks in Sec.~\ref{Sec_remarks}. In the Appendix, we provide an explanation for the origin of
a divergent result encountered in some numerical results. We adopt natural units such that $c=G=\hbar=1$ and the metric signature ($+,-,-,-$).

\section{Electromagnetic field in Schwarzschild spacetime}
\label{field_quantization}
We work with the standard Schwarzschild coordinate system with the line element given by
\begin{equation}
    d\tau^2 = f(r)dt^2 - \frac{dr^2}{f(r)} - r^2(d\theta^2+\sin^2\theta\,d\phi^2),
\end{equation}
where the Schwarzschild radial 
function is
\begin{equation}
\label{Swarzschild}
f(r) = 1-\frac{r_h}{r},
\end{equation}
with $r_h \equiv 2M$ being the radial coordinate of the event horizon.
The dynamics of the electromagnetic field in a modified Feynman gauge can be derived from the following action:
\begin{equation}
\label{action_0}
S = \int \mathcal{L}_{\mathrm{FG}} d^4x,
\end{equation}
with the Lagrangian density given by
\begin{equation}
\label{lagrangian}
\mathcal{L}_{\mathrm{FG}}=\sqrt{-g}\left(-\frac{1}{4}F^{\mu \nu}F_{\mu \nu} - \frac{1}{2}\mathfrak{G}^2\right),
\end{equation}
where
\begin{equation}
  \label{Fmunu}
F_{\mu\nu} = \nabla_{\mu}A_{\nu}-\nabla_{\nu}A_{\mu},
\end{equation}
and
\begin{equation}
\label{H}
\mathfrak{G}\equiv\nabla^{\sigma}A_{\sigma} + K^{\sigma}A_{\sigma}.
\end{equation}
The vector $K^\sigma$ points in the $r$-direction with $K^r = f'(r)$. This choice of $K^{\sigma}$ will prove advantageous as it results in the decoupling of the equation for $A_t$ from the other equations of motion.

The Euler--Lagrange equations are given by
\begin{equation}
\label{Eq_of_motion}
\nabla_{\mu}F^{\mu \nu} + g^{\mu \nu} \nabla_{\mu} \mathfrak{G} - K^{\nu} \mathfrak{G} = 0,
\end{equation}
with (positive-frequency) mode solutions, associated with the timelike Killing vector field $\partial_{t},$ given in the following form:
\begin{equation}
\label{solutions}
A_{\mu}^{\xi n; \omega \ell m} = \eta_{\mu}^{\xi n; \omega \ell m}(r,\theta,\phi)e^{-i\omega t} \hspace{0.5cm} (\omega>0).
\end{equation}
In this equation, the indices $\ell$ and $m$ are the angular quantum numbers; the label $n$ distinguishes between the two kinds of modes, namely, the modes purely incoming from the past null infinity $\mathscr{I}^{-}$ ($n=\mathrm{in}$) and the modes purely incoming from the past (white hole) horizon $H^{-}$ ($n=\mathrm{up}$);
and the index $\xi$ stands for the mode polarization. 

The possible polarizations are summarized as follows:
\begin{equation}
\label{kinds_of_solutions}
\xi \equiv  \begin{cases} \mathfrak{G} &\to \text{pure gauge,} \\
\left.\begin{matrix}
I\\
I\!I
\end{matrix}\ \right\} &\to \text{physical,} \\
N\!P &\to \text{nonphysical.}
\end{cases}
\end{equation}
The pure-gauge polarization gives rise to nonphysical states in the Fock space, which are removed by a Gupta--Bleuler-type physical state condition. The nonphysical polarization gives rise to states with zero norm. (See Ref.~\cite{crispino_2001} for technical details.) Thus, the photon modes other than the physical ones do not influence the observable part of the theory, so that the representative Fock space elements are associated only with the physical modes. Although photon polarizations in curved spacetime have no direct relationship with those in Minkowski spacetime,  it is interesting that in the latter case, the so-called scalar and longitudinal polarizations play a key role in intermediate states (as opposed to asymptotic states).
For example, the Coulomb interaction is envisioned to occur by the exchange of those ``pseudophotons''~\cite{greiner_1996}. For a more detailed discussion about each kind of polarization given by Eq.~\eqref{kinds_of_solutions}, see Refs.~\cite{crispino_2001,castineiras_2005,gubser_1997}. 

From now on, we restrict ourselves to the physical modes  $\xi=I,\,I\!I,$ which satisfy the gauge condition $\mathfrak{G} =0$ for $\ell\geqslant 1$ and give rise to physical states in the Fock space. (The modes with $\ell=0$ are pure gauge or nonphysical.) These modes are explicitly given, in the notation $A_\mu=(A_t,A_r,A_\theta,A_\phi)$, by~\cite{crispino_1998,castineiras_2005}
\begin{eqnarray}
\label{A_I}
A_{\mu}^{I n; \omega \ell m} &=& \bigg(0, \frac{\overline{\varphi_{\omega \ell}^{I n}}}{r^2} Y_{\ell m}, \frac{f(r)}{\ell(\ell+1)} \frac{d \overline{\varphi_{\omega \ell}^{I n}}}{dr} \partial_{\theta} Y_{\ell m}, \nonumber \\
& & \hspace{1cm} \frac{f(r)}{\ell(\ell+1)} \frac{d\overline{\varphi_{\omega \ell}^{I n}}}{dr} \partial_{\phi}Y_{\ell m} \bigg)e^{- i \omega t}, \\
A_{\mu}^{I\!I n; \omega \ell m} &=& \left(0,0, \overline{\varphi_{\omega \ell}^{I\!I n}} Y_{\theta}^{\ell m} , \overline{\varphi_{\omega \ell}^{I\!I n}} Y_{\phi}^{\ell m} \right)e^{- i \omega t},\label{A_II} 
\end{eqnarray}
where the functions $\varphi_{\omega \ell}^{\xi n}(r)$ obey the following differential equation:
\begin{equation}
\label{dif_equation_varphi}
f(r)\frac{d}{dr}\left(f(r)\frac{d}{dr}\varphi_{\omega \ell}^{\xi n}(r)\right) + \left(\omega^2 - V_{\mathrm{eff}}(r)\right)\varphi_{\omega \ell}^{\xi n}(r) = 0,
\end{equation}
with
\begin{equation}
\label{potential}
V_{\mathrm{eff}}(r) \equiv f(r)\frac{\ell(\ell+1)}{r^2}.
\end{equation}
The functions $Y_{\ell m} = Y_{\ell m}(\theta, \phi)$ and $Y^{\ell m}_{\Omega} = Y^{\ell m}_{\Omega}(\theta, \phi)$, $\Omega=\theta,\phi$, are the scalar and vector spherical harmonics, respectively~\cite{NIST_handbook,higuchi_1987b}.
The complex conjugation of the radial modes $\varphi_{\omega \ell}^{\xi n}(r)$ in Eqs.~\eqref{A_I} and \eqref{A_II}, denoted by an overline, converts the $\mathrm{in}$-modes and $\mathrm{up}$-modes to the modes purely outgoing to the future null infinity $\mathscr{I}^{+}$ and those purely ingoing into the future event horizon $H^{+},$ respectively, which are the relevant modes in analyzing the radiation emission rather than the original
$\mathrm{in}$- and $\mathrm{up}$-modes. We use the same labels, ``$\mathrm{in}$'' and ``$\mathrm{up}$'', to indicate these modes and associated quantities.  
We note that only the physical modes labeled by $I$ have a nonzero component in the radial direction. This means that only these modes
contribute to the radiation from the radially infalling charge, as we will see.

The effective potential~\eqref{potential} vanishes asymptotically at the horizon and spatial infinity. Hence, there are analytic solutions satisfying Eq.~\eqref{dif_equation_varphi} such that
\begin{equation}
		\label{asymptotic_in}
\varphi^{\xi \mathrm{in}}_{\omega \ell} = B_{\omega \ell}^{\xi \mathrm{in}} \begin{cases}
 \overline{g(r)} + \mathcal{R}^{\xi \mathrm{in}}_{\omega \ell}g(r), & \hspace{0.2 cm} x \to +\infty, \\
\mathcal{T}^{\xi \mathrm{in}}_{\omega \ell}h(r), & \hspace{0.2 cm} x \to - \infty,
\end{cases}
		\end{equation}
		\begin{equation}
		\label{asymptotic_up}
\varphi^{\xi \mathrm{up}}_{\omega \ell} = B^{\xi \mathrm{up}}_{\omega \ell} \begin{cases}
\overline{h(r)} + \mathcal{R}^{\xi \mathrm{up}}_{\omega \ell}h(r), & \hspace{0.2 cm} x \to - \infty, \\
 \mathcal{T}^{\xi \mathrm{up}}_{\omega \ell}g(r), & \hspace{0.2 cm} x \to + \infty,
\end{cases}
		\end{equation}
where $B^{\xi n}_{\omega \ell}$ are overall normalization constants, and $\mathcal{T}^{\xi n}_{\omega \ell}$ and $\mathcal{R}^{\xi n}_{\omega \ell}$ are the transmission and reflection amplitudes, respectively. The tortoise coordinate $x$ is
defined by $x \equiv r + 2M \ln \left(r/2M - 1\right)$. The complex functions 
$g(r) = e^{i\omega x}[1+O(1/r)]$ and $h(r) = e^{-i\omega x}[1 + O(r-r_h)]$ are expanded as follows:
\begin{eqnarray}
\label{eq:g}
g(r) &=& e^{i \omega x}\sum_{j=0}^{j_{\mathrm{max}}} \frac{g_j}{r^j},\\
\label{eq:h}
h(r) &=& e^{-i \omega x}\sum_{j=0}^{j_{\mathrm{max}}} h_j (r-r_h)^{j},
\end{eqnarray}
where $h_j$ and $g_j$ are complex coefficients obtained by solving Eq.~\eqref{dif_equation_varphi} order by order near the horizon and infinity (see, e.g., Ref.~\cite{pani_2013}) starting from $g_0=h_0=1$.
The order of the expansion is associated with the choice of $j_{\mathrm{max}}.$ We choose $j_{\mathrm{max}}=20$ in our numerical computation.

The solutions $\varphi^{\xi n}_{\omega \ell}$ to Eq.~\eqref{dif_equation_varphi} satisfy the boundary conditions specified in Eqs.~\eqref{asymptotic_in} and~\eqref{asymptotic_up}. By matching the numerical solution for $\varphi^{\xi n}_{\omega \ell}$ with the boundary conditions, we determine the coefficients $\mathcal{T}^{\xi n}_{\omega \ell}$ and $\mathcal{R}^{\xi n}_{\omega \ell}$. 
As is well known, these coefficients are not independent. By using the properties of the Wronskian of the asymptotic solutions, given by Eqs.~\eqref{asymptotic_in} and~\eqref{asymptotic_up}, we obtain the conservation relation,
\begin{equation}
\label{flux_conservation}
\abs{\mathcal{T}^{\xi n}_{\omega \ell}}^2 + \abs{\mathcal{R}^{\xi n}_{\omega \ell}}^2 = 1. 
\end{equation} 

The conserved current $W^{\mu}$ associated to two solutions $A^{(i)}_\mu$ and $A^{(j)}_\mu$ is defined as
\begin{equation}
\label{current}
W^{\mu}[A^{(i)},A^{(j)}] = i \left[\overline{A_{\sigma}^{(i)}} \Pi^{\mu \sigma}_{(j)} - \overline{\Pi^{\mu \sigma}_{(i)}}A_{\sigma}^{(j)}\right],
\end{equation}
where the canonical conjugate momentum current
$\Pi_{(i)}^{\mu \nu},$ associated with the solution $A_{\nu}^{(i)},$ is defined by
\begin{eqnarray}
\label{conjugate_momenta}
\Pi_{(i)}^{\mu \nu} &\equiv & \frac{1}{\sqrt{-g}}\frac{\partial \mathcal{L}}{\partial[\nabla_{\mu}A_{\nu}]}\Big\vert_{A_{\mu} = A_{\mu}^{(i)}} \nonumber\\
& = & \left[- F^{\mu \nu} - g^{\mu \nu} \mathfrak{G} \right]\big\vert_{A_{\mu} = A_{\mu}^{(i)}}.
\end{eqnarray}
The generalized Klein--Gordon inner product, given by
\begin{equation}
\label{KG}
\left(A^{(i)},A^{(j)} \right) = \int_{\Sigma} d\Sigma\, n_{\mu} W^{\mu}[A^{(i)},A^{(j)}],
\end{equation}
is used to normalize the modes,
where $\Sigma$ is a Cauchy hypersurface for the
exterior region of the Schwarzschild spacetime, with $n^{\mu}$ being the future-pointing unit normal to $\Sigma$.
For the physical modes with $\xi = I,\,I\!I,$ we impose the orthogonality relation,
\begin{equation}
\label{ortogonality_relation}
\left(A^{\xi n; \omega \ell m},A^{\xi' n'; \omega' \ell' m'} \right) = \delta_{\xi \xi'}\delta_{n n'}\delta_{\ell \ell'}\delta_{m m'} \delta(\omega - \omega'),
\end{equation}
where $(i)$ and $(j)$ in Eqs.~\eqref{current}--\eqref{KG} represent
the set of labels $\xi n; \omega \ell m.$ 
From Eqs.~\eqref{asymptotic_in},~\eqref{asymptotic_up},~\eqref{KG}, and~\eqref{ortogonality_relation}, the overall normalization constants for $\xi=I,\,I\!I$ are readily obtained as 
\begin{equation}
\label{overall_normalization}
\abs{B^{I n}_{\omega \ell}} = \sqrt{\frac{\ell(\ell+1)}{4\pi \omega^3}}, \hspace{1. cm} \abs{B^{I\!I n}_{\omega \ell}} = \frac{1}{\sqrt{4\pi\omega}}.
\end{equation}

The quantum field operator $\hat{A}_{\mu}$ corresponding to the classical field $A_\mu$
 is expanded in terms of positive and negative frequency modes,
\begin{equation}
\label{field_operator}
\hat{A}_{\mu} = \sum_{\xi, n, \ell, m} \int_{0}^{\infty}d\omega \left[\hat{a}_{(i)} A_{\mu}^{(i)} + \hat{a}_{(i)}^{\dagger} \overline{A_{\mu}^{(i)}} \right].
\end{equation}
After imposing the standard equal-time commutation relations on the quantum field 
operators $\hat{A}_{\mu}$ and $\hat{\Pi}^{t \nu}$ corresponding to the fields $A_\mu$ and $\Pi^{t\nu}$, respectively, 
one finds that the annihilation and creation operators, $\hat{a}_{(i)}$ and $\hat{a}_{(i)}^{\dagger}$, have the following nonvanishing commutation relations for physical modes:
\begin{equation}
\label{commutation}
\left[\hat{a}_{\xi n; \omega \ell m}, \hat{a}_{\xi' n'; \omega' \ell' m'}^{\dagger} \right] = \delta_{\xi \xi'}\delta_{n n'}\delta_{\ell \ell'}\delta_{m m'} \delta(\omega - \omega'),
\end{equation}
where $\xi, \xi' = I, I\!I$.

We follow the Gupta--Bleuler quantization prescription and find that the physically relevant states are represented in the Fock space by 
those obtained applying the creation operators associated with the physical modes $\xi=I,\,I\!I$ to the Boulware vacuum state $\ket{0}$, defined by 
$\hat{a}_{\xi n; \omega \ell m}\ket{0} = 0$~\cite{boulware_1975}, 
in the sense that any
physically relevant state differs from one of these states by a zero-norm state. In particular, the (representative) one-particle  states are given by
\begin{equation}
\label{one_particle_state}
 \ket{\xi n; \omega \ell m} = \hat{a}_{\xi n; \omega \ell m}^{\dagger} \ket{0}.
\end{equation}

In the next section, we analyze the interaction between a classical charged particle falling into the Schwarzschild BH and the quantum electromagnetic field $\hat{A}_\mu.$ In particular, we find the probability of the charged particle emitting one photon, which corresponds to the emission of radiation in classical electrodynamics.
\section{Radiation emission}
\label{radiation_emission}

\subsection{Infalling charged particle}
\label{geodesics}
The electrically charged point particle is described by the following current density:
\begin{equation}
\label{source}
j^{\mu}(x) = \frac{q\,v^{\mu}}{\sqrt{-g} v^t}\delta(r-r_{s})\delta(\theta - \theta_{s})\delta(\phi-\phi_{s}),
\end{equation}
where $r_s$, $\theta_s,$ and $\phi_s$ are the spatial coordinates of the particle and where $v^{\mu}$ is its $4$-velocity,
\begin{equation}
\label{four_velocity}
v^{\mu} = \frac{dx^{\mu}}{d\tau} = \left(\frac{E}{f(r)}, - \sqrt{E^2 - f(r)}, 0, 0 \right),
\end{equation}
with $\tau$ being the proper time of the particle.
In Eq.~\eqref{four_velocity}, the
quantity $E$ is the specific energy, i.e., the energy per unit rest mass of the particle,
as inferred by the inertial observer $\mathcal{O}$ far away from the BH, and it can be given in terms of the initial radial position and velocity, $r_{0}$ and $v_{0}$, as
\begin{equation}
\label{energy}
E = \sqrt{\frac{f(r_0)}{1-v_0^2/f(r_0)^2}}.
\end{equation}
The radial velocity of the particle, as seen by 
the observer $\mathcal{O},$ is given by
\begin{equation}
\label{reciprocal_radial_velocity}
v_s \equiv - \frac{dr_s}{dt} =  \frac{f(r_s)\sqrt{E^2 - f(r_s)}}{E}.
\end{equation}
According to the observer $\mathcal{O},$ the particle experiences acceleration and/or deceleration, depending on the initial conditions. For sufficiently small $v_0,$ the velocity $v_s$ increases from $v_0,$ reaches a maximum at $r_s=6M/(3 - 2E^2)$ (at $r_s = 6M$ for $v_0=0$ with $r_0=\infty$), and then decreases to zero at the horizon. However, if the condition $v_0> \sqrt{(16M^3-12M^2r_0+r_0^3)/3r_0^3}$ ($v_0> 1/\sqrt{3}$ for $r_0 = \infty$) holds, $v_s$ has no maximum, and the particle only decelerates when projected from $r_s=r_0.$
We also observe that, for $v_0=0$, the maximum 
acquired radial velocity decreases with decreasing $r_0.$ For $r_0 \to \infty$ and $v_0=0,$ the maximum acquired radial velocity is $2/(3\sqrt{3})\approx 0.38.$ For a
static observer very close to the horizon, the charge always passes by them with the radial velocity close to $1$.

In the next subsection, we use QFT at tree level to obtain the one-particle-emission amplitude.

\subsection{One-particle-emission amplitude}
\label{sub_sec_one_particle_emission}
The coupling of the classical charge to the quantum field is given by the interaction action,
\begin{equation}
\label{int_lagrangian}
\hat{S}_{\mathrm{int}} = \int \sqrt{-g} j^{\mu}\hat{A}_{\mu}\,d^4x.
\end{equation}
The current density $4$-vector $j^{\mu}$ given by Eq.~\eqref{source} is conserved,
i.e., $\nabla_{\mu}j^{\mu}=0.$ For any Cauchy hypersurface $\Sigma,$ we have $\int_{\Sigma}d\Sigma\,n_\mu j^{\mu} = q$.

The interaction action given by Eq.~\eqref{int_lagrangian} gives rise to a nonvanishing probability amplitude at first order. For the emission of a physical photon with polarization $\xi,$ energy $\omega,$ and angular quantum numbers $\ell$ and $m,$ it is given by
\begin{eqnarray}
\label{one_particle_emission}
\mathcal{A}^{\xi n; \omega \ell m} &=& \bra{\xi n; \omega \ell m} i \hat{S}_{\mathrm{int}} \ket{0} \\
\label{one_particle_emission_2}
&=& i \int \sqrt{-g} j^{\mu} \overline{A_{\mu}^{\xi n; \omega \ell m}}\,d^4x.
\end{eqnarray}
As mentioned earlier, since the charged particle is falling radially, the current density $j^{\mu}$ only couples to the modes with
$A_{r}^{\xi n; \omega \ell m} \neq 0$, among the
physical modes [see Eqs.~\eqref{source} and~\eqref{four_velocity}]. Therefore, the modes with $\xi=I\!I$ are not excited by the falling charge.\footnote{The coupling to the modes depends on the motion of the charged particle. For example, a charge orbiting the BH along a circular geodesic~\cite{castineiras_2005} or plunging into the BH due to a perturbation in its unstable circular orbit are examples in which both the modes with $\xi=I$ and $\xi=I\!I$ are excited.} From now on, we omit the index $\xi$ with the understanding that $\xi = I$.

Two alternative initial states to the Boulware vacuum state used in Eq.~\eqref{one_particle_emission} are the Unruh vacuum state~\cite{unruh_1976}, characterized by an outward thermal flux at future null infinity and no incoming flux at past null infinity, and the Hartle--Hawking vacuum state~\cite{hartle_1976}, characterized by thermal fluxes across both past and future null infinities. In these cases, we would have to take into account absorption and stimulated emission of photons induced by the thermal fluxes. This would lead to additional transition amplitudes to be calculated, where Bose--Einstein thermal factors play a part. However, one can show that the absorption and stimulated emission amplitudes exactly cancel, and the resulting {\it net} radiation is the same as that calculated using the Boulware vacuum state.

Substituting Eqs.~\eqref{A_I} and~\eqref{source} into Eq.~\eqref{one_particle_emission_2}, we obtain
\begin{equation}
\label{amplitude_parcial}
\mathcal{A}^{n; \omega \ell m} = i q \overline{Y_{\ell m}} \int_{- \infty}^{+\infty}dt_s \frac{v^r}{v^t} \frac{\varphi_{\omega \ell}^{n}}{r_s^2}e^{i \omega t_s} ,
\end{equation}
with $r_s = r(t_s)$ denoting the position of the charged particle at $t=t_s.$ 
It is not possible to find a closed-form expression for
$\mathcal{A}^{n; \omega \ell m}$ in Eq.~\eqref{amplitude_parcial}, for an arbitrary value of
$\omega$, but it is possible to find
an analytic expression for it in the $\omega \to 0$ limit, as we will see.

We consider the charged particle static at $r=r_0$, for $-\infty < t < 0$, and projected toward the BH at $t=0.$
(Note that the function $r(t_s)$ is one-to-one for $t_s \in [0, \infty).$)
By integrating by parts and changing the integration
variable from $t_s$ to $r_s$, we find
\begin{equation}
\label{amplitude_Perman}
\mathcal{A}^{n; \omega \ell m} = \frac{q \overline{Y_{\ell m}}}{\omega} \int_{2M}^{r_0} \left. \dfrac{d}{dr} \left(\frac{v^r}{v^t} \frac{\varphi_{\omega \ell}^{n}}{r^2} \right)\right|_{r=r_s} e^{i \omega t(r_s)} dr_s.
\end{equation}
Note that the spatial components of the current, corresponding to the charge at rest, vanish, and hence, this charge does not couple to the physical modes given by Eqs.~\eqref{A_I} and~\eqref{A_II}. 
Using the delta function identity~\cite{NIST_handbook},
\begin{equation}
\label{delta_property}
\delta(r(t)-r(t_s)) = \frac{v^t}{\abs{v^r}}\delta(t-t_s),
\end{equation}
and Eq.~\eqref{reciprocal_radial_velocity}, we can rewrite the radial component of the current density given by Eq.~\eqref{source} as
\begin{equation}
\label{current_with_t}
j^{r}(x) = - \frac{q}{\sqrt{-g}} \delta(t-t_s)\delta(\theta - \theta_s) \delta(\phi-\phi_s).
\end{equation}
Substituting Eq.~\eqref{current_with_t} into Eq.~\eqref{one_particle_emission_2} [or changing
the integration variable from $t_s$ to $r_s$ in Eq.~\eqref{amplitude_parcial}], we obtain
\begin{equation}
\label{amplitude_sudd}
\mathcal{A}^{n; \omega \ell m} = -i q \overline{Y_{\ell m}} \int_{2M}^{r_0} \frac{\varphi_{\omega \ell}^{n}(r_s)}{r_s^2} e^{i \omega t(r_s)} dr_s.
\end{equation}
It would appear that the two expressions of the
emission amplitude, Eqs.~\eqref{amplitude_Perman} and \eqref{amplitude_sudd}, 
differ by the following boundary term:
\begin{equation}
\label{boundary_term}
\mathcal{A}^{n; \omega \ell m}_{\mathrm{boundary}}
= - q \overline{Y_{\ell m}} \left. v_s \frac{\varphi_{\omega \ell}^{n}(r_s)}{\omega r_s^2} \right|_{r_s=r_0}.
\end{equation} 
This boundary term is proportional to $v_0$ and $r_0^{-2},$ implying that the two expressions of the emission amplitude
coincide, if $v_0 = 0$ or $r_0 \to \infty.$\footnote{The factor $v^r$ in the boundary term, given by Eq.~\eqref{boundary_term}, does not appear in the scalar field case (see Eq. (36) of Ref.~\cite{oliveira_2018} and note that $v_s=-v^r/v^t$). In the scalar case, the boundary term
corresponding to Eq.~\eqref{boundary_term} vanishes only for $r_0\to \infty.$}
Although one can argue that the boundary term~\eqref{boundary_term} should be absent and that Eq.~\eqref{amplitude_Perman} should be adopted even if $v_0\neq 0$ and $r_0 < \infty$, this case represents a
point charge with infinite acceleration at $t=0$, which is unphysical.
Therefore, we specialize to the cases with $v_0=0$ or $r_0\to \infty$, for which Eqs.~\eqref{amplitude_Perman} and \eqref{amplitude_sudd} coincide.

Using the one-particle-emission amplitude, we can derive the partial energy spectrum, which
refers to the energy spectrum for each multipole $\ell$,
\begin{equation}
\label{energy_spectrum_formula}
\mathcal{E}^{n; \omega \ell} = \sum_{m=-\ell}^{\ell} \omega \abs{\mathcal{A}^{n; \omega \ell m}}^2.
\end{equation}
We sum over $m$ by using the formula
\begin{equation}
\label{m_contribution}
 \sum_{m=-\ell}^{\ell} \overline{Y_{\ell m}(\theta_s, \phi_s)}Y_{\ell m}(\theta_s, \phi_s) = \frac{2\ell+1}{4\pi},
 \end{equation}
 and find
 \begin{equation}
 \label{energy_spectrum_sudd}
 \mathcal{E}^{n ; \omega \ell} = \frac{(2\ell+1)q^2 \omega}{4\pi} \abs{\int_{2M}^{r_0} \frac{\varphi_{\omega \ell}^{n}(r_s)}{r_s^2}e^{i\omega t(r_s)}dr_s}^2.
 \end{equation}
Integrating Eq.~\eqref{energy_spectrum_sudd} over $\omega > 0,$ we obtain the partial emitted energy, i.e., the emitted energy 
associated with each multipole $\ell$,
\begin{equation}
\label{partial_energy}
\mathcal{E}^{n ; \ell} = \int_{0}^{\infty} d\omega\, \mathcal{E}^{n ; \omega \ell}.
\end{equation}
We also calculate the total energy spectrum $\mathcal{E}^{n ; \omega}$ by summing the contributions of all multipoles in Eq.~\eqref{energy_spectrum_sudd}: dipole $(\ell = 1),$ quadrupole $(\ell=2),$ octupole $(\ell=3),$ hexadecapole $(\ell=4),$ and so on. Thus,
\begin{equation}
\label{total_energy_spectrum}
\mathcal{E}^{n ; \omega} = \sum_{\ell \geqslant 1} \mathcal{E}^{n ; \omega \ell}.
\end{equation}
The total emitted energy, obtained from Eq.~\eqref{energy_spectrum_sudd}, is given by
\begin{equation}
\label{total_energy}
\mathcal{E}^{n} = \sum_{\ell \geqslant 1} \int_{0}^{\infty} d\omega\, \mathcal{E}^{n ; \omega \ell}.
\end{equation}

The energy emitted to infinity is associated with the time-reversed $\mathrm{in}$-modes, while the energy absorbed by
the BH is associated with the time-reversed
$\mathrm{up}$-modes.
The $\mathrm{in}$-modes are purely incoming from the past null infinity $\mathscr{I}^-.$ Hence, the time-reversed $\mathrm{in}$-modes are purely outgoing to the future null infinity $\mathscr{I}^+$, and the $\mathrm{up}$-modes are purely incoming from the past event horizon $H^{-}.$ Hence, the time-reversed $\mathrm{up}$-modes are purely outgoing into the future event horizon $H^+$ (see, e.g., Ref.~\cite{crispino_2000}). The time reversal is achieved by the complex conjugation in Eqs.~\eqref{A_I} and \eqref{A_II}, as we stated before.

The amplitude $\mathcal{A}^{n ; \omega \ell m}$ is determined by carrying out the integral in Eq.~\eqref{amplitude_sudd} numerically, where $\varphi_{\omega \ell}^{n}$ is obtained by numerically solving the differential Eq.~\eqref{dif_equation_varphi}, with boundary conditions given by Eqs.~\eqref{asymptotic_in} and~\eqref{asymptotic_up}. Before presenting our numerical results in
Sec.~\ref{sec_results}, we compute analytically the emission amplitude and the corresponding partial energy spectra in the zero-frequency limit~\cite{higuchi_1997,crispino_1998,oliveira_2018}
in the next section. 

\section{zero-frequency limit}
\label{low_frequency}
In this section, we find analytically the zero-frequency limit of some quantities we defined in the previous section.  
Comparison of these quantities with the corresponding numerical results serves as
a consistency check for the numerical method. We verify that the numerical method and these analytical quantities are in very good agreement with each other.

\subsection{$\mathbf{In}$-modes for $r_0 \to \infty$}
\label{Inmoder0infty}
In this subsection, we study the $\mathrm{in}$-mode solutions in the zero-frequency limit for $r_0\to\infty$ and arbitrary $v_0.$ 
Adapting the method used in Ref.~\cite{oliveira_2018}, we can obtain the emission amplitude associated with the radiation emitted to infinity in the low-frequency regime. In this regime, one has $M \omega \ll 1,$ which is equivalent to letting $f(r) \approx 1,$
and hence, $t(r_s) = -v_0^{-1} r_s.$ (The time coordinate $t$, when the charged particle is released, is not $0$, but this does not affect the result.) Thus, the zero-frequency limit of the $\mathrm{in}$-modes coincides with the flat-spacetime limit.
Therefore, we have in this limit,
\begin{equation}
\label{low_freq_phi_1}
\varphi_{\omega \ell}^{\mathrm{in}} = C_{\omega} \sqrt{\frac{2 \omega}{\pi}}\,r j_{\ell}(r \omega),
\end{equation}
where $C_{\omega}$ is a normalization constant, and $j_{\ell}(y)$ is the spherical Bessel function of order $\ell$. Comparing Eq.~\eqref{low_freq_phi_1} with Eq.~\eqref{asymptotic_in}, we obtain $C_{\omega} =\sqrt{2 \pi \omega}B_{\omega \ell}^{I \mathrm{in}}$ (up to a phase factor).  
Using Eqs.~\eqref{low_freq_phi_1} and~\eqref{amplitude_sudd}, we obtain 
\begin{equation}
\label{low_freq_ampli_sudd}
\mathcal{A}^{\mathrm{in}; 0 \ell m} = 2 iq    \sqrt{\frac{\ell(\ell+1)}{4\pi \omega}}\overline{Y_{\ell m}} \int_{0}^{\infty} \frac{j_{\ell}(r_s \omega)}{r_s}e^{-i \omega r_s/v_0} dr_s.
\end{equation}
Notice that the integral is $\omega$ independent.
The associated partial energy spectrum for $\ell\geqslant 1$ is given, after evaluating the integral in this equation~\cite[Eqs.~6.699.1,2]{gradshteyn}, by
\begin{eqnarray}
\label{low_freq_partial_energy}
\mathcal{E}^{\mathrm{in} ; 0 \ell} &=& q^2 \frac{(2\ell+1)\ell(\ell+1)\Gamma(\ell)^2}{16 \pi\cdot 4^{\ell} \Gamma(\ell+\frac{3}{2})^2} v_0^{2 \ell}  \nonumber \\  && \hspace{0.75cm} \times \abs{_2F_1 \left(\frac{\ell}{2}, \frac{\ell + 1}{2}; \ell +\frac{3}{2}; v_0^2 \right)}^2,
\end{eqnarray}
where $_2F_1$ is the 
Gauss hypergeometric function. We note that in the limit $v_0 \to 0$ the energy given by Eq.~\eqref{low_freq_partial_energy} vanishes
like $v_0^{2\ell};$ i.e., for small values of $v_0,$ we have
\begin{equation}
\label{low_freq_partial_energy_limit_v_0_0}
\mathcal{E}^{\mathrm{in} ; 0 \ell} \approx q^2  \frac{(2\ell+1)\ell(\ell+1)\Gamma(\ell)^2}{16 \pi\cdot 4^{\ell} \Gamma(\ell+\frac{3}{2})^2}  v_0^{2 \ell}.
\end{equation}
On the other hand, in the limit $v_0 \to 1,$ we find
the following expression from Eq.~\eqref{low_freq_partial_energy}, using Ref.~\cite[Eq.~9.122.1]{gradshteyn}:
\begin{equation}
\label{low_freq_partial_energy_limit_v_0_1}
\lim_{v_0 \to 1} \mathcal{E}^{\mathrm{in} ; 0 \ell} \equiv \mathcal{E}^{\mathrm{in} ; \omega \ell}_{\mathrm{class}} = \frac{q^2}{4\pi^2}\frac{2\ell+1}{\ell(\ell+1)}.
\end{equation}
This is exactly the $\omega$-independent classical result obtained in electromagnetism in flat spacetime 
when a charged particle is suddenly decelerated (see, e.g., Ref.~\cite{cardoso_eletr_2003}).\footnote{Note that we are using rationalized units (see, e.g., the Appendix in Ref~\cite{jackson}).} 

Let us digress here and discuss the energy spectra for the case $v_0\to 1$ as a whole, including features not necessarily related to the low-frequency 
limit.
The $\ell$-sum of Eq.~\eqref{low_freq_partial_energy_limit_v_0_1} gives a divergent result for the total
emitted energy spectrum $\mathcal{E}^{\mathrm{in};\omega}_{\mathrm{class}}$ for flat spacetime. This indicates that 
the total energy spectrum $\mathcal{E}^{\mathrm{in};\omega}$ for the BH is also divergent for $v_0\to 1$~\cite{cardoso_eletr_2003}.

The flat-spacetime counterpart of the partial energy $\mathcal{E}^{\mathrm{in} ; \ell}$, emitted by the particle, 
diverges in the limit $v_0\to 1$, because the partial spectrum $\mathcal{E}^{\mathrm{in} ; \omega \ell}_{\mathrm{class}}$ is $\omega$ independent.
For the BH, however, the partial energy $\mathcal{E}^{\mathrm{in} ; \ell}$ 
can be estimated by introducing a cutoff frequency  in Eq.~\eqref{low_freq_partial_energy_limit_v_0_1}, which we choose to be the associated fundamental quasinormal frequency $\omega^{\mathrm{qnf}}_\ell$. (This choice is motivated by our numerical results, which reveal that $\mathcal{E}^{\mathrm{in} ;\omega \ell}$ decays exponentially for $\omega > \omega_{\ell}^{\mathrm{qnf}}$, as shown in Fig.~\ref{to_infinity_r0_inf_ultrarelat}.)
We can approximate the frequencies $\omega^{\mathrm{qnf}}_\ell$ as $\omega^{\mathrm{qnf}}_\ell \approx \sqrt{\ell(\ell+1)}/b_c,$ where $b_c=3\sqrt{3}M$ is the critical impact parameter of null geodesics~\cite{goebel_1972}.
With this approximation, we find
\begin{equation}
\label{classical_partial_energy}
\mathcal{E}^{\mathrm{in} ; \ell}
\approx \mathcal{E}^{\mathrm{in} ; \omega \ell}_{\mathrm{class}} \omega_\ell^{\mathrm{qnf}} \approx \frac{q^2}{4\pi^2}\frac{2}{b_c}\,\,\textrm{for}\ v_0 \approx 1.
\end{equation}
For a given value of $v_0$ very close to $1$, the first
approximate equality in Eq.~\eqref{classical_partial_energy} becomes more accurate if we use Eq.~\eqref{low_freq_partial_energy} instead of Eq.~\eqref{low_freq_partial_energy_limit_v_0_1}: the spectrum found numerically is nearly $\omega$ independent up to $\omega \approx \omega_{\ell}^{\mathrm{qnf}}$, for charges in ultrarelativistic motion. Thus,
\begin{equation}
\label{partial_energy_v0_approx_1}
\mathcal{E}^{\mathrm{in} ; \ell}
\approx \mathcal{E}^{\mathrm{in} ; 0 \ell} \omega_\ell^{\mathrm{qnf}}\ \textrm{for}\ v_0 \approx 1,
\end{equation}
where $\mathcal{E}^{\mathrm{in} ; 0 \ell}$ is given by Eq.~\eqref{low_freq_partial_energy}. (See Refs.~\cite{smarr_1977,cardoso_2002} for a similar discussion within the framework of gravitational radiation.)

Table~\ref{energy_spectra_infinty_comparison_sudd} shows a comparison of the zero-frequency limit
of the partial energy spectrum obtained through Eq.~\eqref{low_freq_partial_energy} and this quantity obtained by solving Eq.~\eqref{dif_equation_varphi} and carrying out the integral in Eq.~\eqref{amplitude_sudd} numerically.
We see that our numerical computations and the analytical zero-frequency limit are in very good agreement. 
		\begin{table}[h!]
		\begin{center}
\begin{tabular}{||c | c | c  c||} 
 \hline
 $\ell$ & $v_0$ & \hspace{0.5cm} $q^{-2}\mathcal{E}^{\mathrm{in} ; 0 \ell}$ \hspace{0.5cm} & \hspace{0.5cm} Numerical \hspace{0.5cm} \\ [0.5ex] 
 \hline\hline
 \multirow{3}{0.7cm}{\hspace{0.2cm} 1} & $0.25$ & $0.0010827$ & $0.0010828$ \\ 

  & $0.75$ & $0.0126318$ & $0.0126317$ \\

  & $0.99$ & $0.0347517$ & $0.0347628$ \\
 \hline
 \multirow{3}{0.7cm}{\hspace{0.2cm} 2} & $0.25$ & $0.0000139$ & $0.0000139$ \\ 

  & $0.75$ & $0.0019812$ & $0.0019813$ \\

  & $0.99$ & $0.0170247$ & $0.0170315$ \\
   \hline
 \multirow{3}{0.7cm}{\hspace{0.2cm} 3} & $0.25$ & $0.0000002$ & $0.0000002$ \\ 

  & $0.75$ & $0.0003610$ & $0.0003611$ \\

  & $0.99$ & $0.0102605$ & $0.0102606$ \\ 
 \hline
		\end{tabular}	
		\caption{Comparison between the analytical results for $\mathcal{E}^{\mathrm{in} ; 0 \ell},$ given by Eq.~\eqref{low_freq_partial_energy}, and the numerically obtained partial energy spectrum with $\omega \to 0,$ for the first three multipoles and representative choices of $v_0.$}
		 \label{energy_spectra_infinty_comparison_sudd}
		\end{center}
		\end{table}

\subsection{$\mathbf{In}$-modes for finite $r_0$}

In this subsection, we study the cases in which $r_0$ is finite. For finite $r_0,$ the emission amplitude vanishes for all $\ell \geqslant 1$ in the zero-frequency limit.
This is verified by noting that the $\omega =0$ solutions to $\varphi_{\omega \ell}^{\mathrm{in}},$ given in Ref.~\cite{castineiras_2005}, are expressed as
\begin{equation}
\label{low_freq_varphi_in}
\varphi_{\omega \ell}^{\mathrm{in}} \approx \mathcal{C}^{\mathrm{in}}_{\omega \ell} r \left[ P_{\ell}\left(r/M-1\right) - \frac{(r - 2M)}{\ell(\ell+1)} \frac{d}{dr}P_{\ell}(r/M -1) \right],
\end{equation}
for $\omega \approx 0$,
where $P_{\ell}$ are Legendre functions of the first kind and
\begin{equation}
\label{constant_to_Pl}
\abs{\mathcal{C}^{\mathrm{in}}_{\omega \ell}} = \frac{1}{\sqrt{\pi \ell(\ell+1)}} \frac{2^\ell ((\ell+1)!)^2 M^\ell}{(2\ell)! (2\ell+1)!!} \omega^{\ell-1/2}.
\end{equation}
We see from Eqs.~\eqref{low_freq_varphi_in} and~\eqref{constant_to_Pl} that $\varphi_{\omega \ell}^{\mathrm{in}} \propto \omega^{\ell-1/2}$ for $\omega \approx 0$ and the quantity $\mathcal{E}^{\mathrm{in} ; \omega \ell}$ vanishes like $\omega^{2\ell}$ as $\omega \to 0.$
\subsection{$\mathbf{Up}$-modes}
\label{up-finite-r0}
In this subsection, we study the $\mathrm{up}$-mode solutions in the zero-frequency limit.   
 Low-frequency solutions for $\varphi_{\omega \ell}^{\mathrm{up}}$ are~\cite{castineiras_2005}
\begin{equation}
\label{low_freq_varphi_up}
\varphi_{\omega \ell}^{\mathrm{up}} \approx \mathcal{C}^{\mathrm{up}}_{\omega \ell} r \left[ Q_{\ell}(r/M-1) - \frac{(r - 2M)}{\ell(\ell+1)} \frac{d}{dr}Q_{\ell}(r/M -1) \right],
\end{equation}
where $Q_{\ell}$ are Legendre functions of the second kind and
\begin{equation}
\label{constant_to_Ql}
\abs{\mathcal{C}^{\mathrm{up}}_{\omega \ell}} =  2 \sqrt{\frac{\ell(\ell+1)}{\pi}}\omega^{-1/2}.
\end{equation}
Therefore, we have $\varphi_{\omega \ell}^{\mathrm{up}} =O(\omega^{-1/2})$ for small $\omega$. Writing the exponential in the integrand of Eq.~\eqref{amplitude_sudd} as an infinite power series
in $\omega$, we see that only the first term of $\sqrt{\omega}\mathcal{A}^{\mathrm{up}; 0 \ell m}$ in this series will be nonzero [and is independent of $v_0$ because the $v_0$ dependence enters only into the function $t(r_s)$]. Using Eq.~\eqref{low_freq_varphi_up}, one can easily obtain the absorbed partial energy spectrum in the zero-frequency limit, $\mathcal{E}^{\mathrm{up} ; 0 \ell}.$ 

\begin{figure}
\center
\includegraphics[scale=0.45]{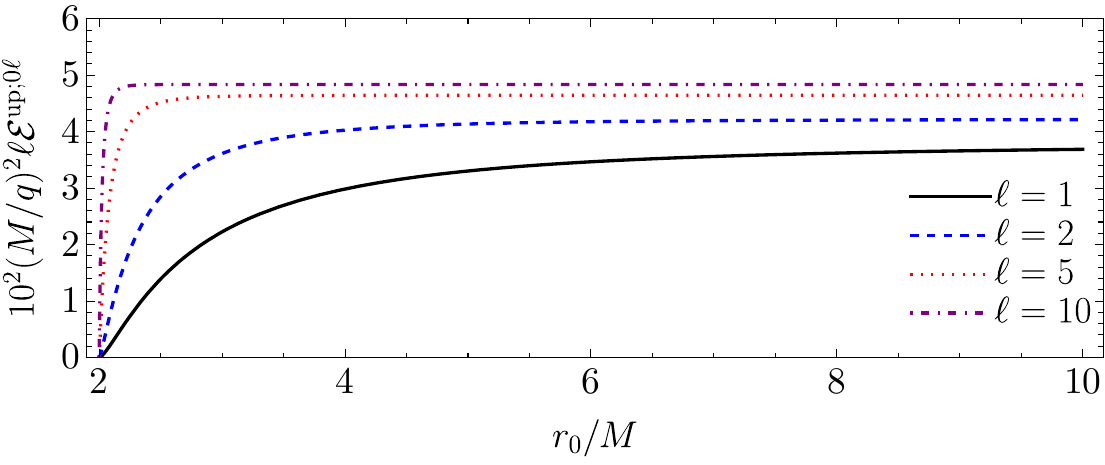}
\caption{The zero-frequency limit of the partial energy spectrum, $\mathcal{E}^{\mathrm{up}; 0 \ell}$, as a function of $r_0$ for some choices of $\ell.$}
\label{Energy_low_omega_r0}
\end{figure}

Table~\ref{energy_spectra_horizon_r0_finite_v0_subto_comparison} shows a comparison of the zero-frequency limit of the energy spectrum, $\mathcal{E}^{\mathrm{up} ; 0 \ell}$, and the corresponding numerical result.
~Figure~\ref{Energy_low_omega_r0} shows the partial energy spectrum $\mathcal{E}^{\mathrm{up} ; 0 \ell}$ as a function of $r_0.$ As $\ell$ increases, the spectrum at the zero-frequency limit converges to a value almost independent of $r_0.$
This is because the function $\varphi^{\mathrm{up}}_{\omega \ell}(r)$ tends to zero like $r^{-\ell}$ for large $r$ at low frequencies [see Eq.~\eqref{low_freq_varphi_up}] in 
the integral~\eqref{amplitude_sudd} for the amplitude.
		\begin{table}[h!]
		\begin{center}
\begin{tabular}{||c | c | c  c||} 
 \hline
 $\ell$ & $r_0$ & \hspace{0.5cm} $(M/q)^{2}\mathcal{E}^{\mathrm{up} ; 0 \ell}$ \hspace{0.5cm} & \hspace{0.5cm} Numerical \hspace{0.5cm} \\ [0.5ex] 
 \hline\hline
 \multirow{3}{0.7cm}{\hspace{0.2cm} 1} & $3M$ & $0.02225212$ & $0.02225193$ \\ 

  & $6M$ & $0.03466686$ & $0.03466683$ \\

  & $100M$ & $0.03798520$ & $0.03798522$ \\
   & $\infty$ & $0.03799544$ & $0.03799548$ \\
 \hline
 \multirow{3}{0.7cm}{\hspace{0.2cm} 2} & $3M$ & $0.01806131$ & $0.01806121$ \\ 

  & $6M$ & $0.02088009$ & $0.02088010$ \\

  & $100M$ & $0.02110854$ & $0.02110856$ \\
     & $\infty$ & $0.02110858$ & $0.02110862$ \\
   \hline
 \multirow{3}{0.7cm}{\hspace{0.2cm} 3} & $3M$ & $0.01410750$ & $0.01410746$ \\ 

  & $6M$ & $0.01475750$ & $0.01475751$ \\

  & $100M$ & $0.01477600$ & $0.01477604$ \\ 
  & $\infty$ & $0.01477601$ & $0.01477605$ \\ 
 \hline
		\end{tabular}	
		\caption{Comparison between the analytical results of the partial energy spectrum $\mathcal{E}^{\mathrm{up} ; \omega \ell}$ and the corresponding numerical results in the $\omega \to 0$ limit, for the first three multipoles and some choices of $r_0.$} \label{energy_spectra_horizon_r0_finite_v0_subto_comparison}
		\end{center}
		\end{table}
		
\section{Numerical results}
\label{sec_results}
In this section, we show some results for nonzero frequencies $ \omega,$ obtained by numerically solving Eq.~\eqref{dif_equation_varphi} from $r=2M(1+\epsilon)$ (with $\epsilon \equiv 10^{-5}$) to the numerical infinity $r_{\infty},$ which we choose to be~\cite{bernar_2017}
\begin{equation}
\label{numerical_infinity}
r_{\infty} \equiv 250 \frac{\sqrt{\ell(\ell+1)}}{\omega}.
\end{equation}
 With these choices, we achieve good precision, evidenced in the previous section from the very good agreement between the numerical and analytical results in the zero-frequency limit.

In this section, the vertical gray lines in the plots correspond to the fundamental quasinormal frequencies $\omega_{\ell}^{\mathrm{qnf}}$ of the BH, and the horizontal gray lines mark the values of the associated zero-frequency limit, unless otherwise stated.

In the next subsection, we analyze the numerical results associated with the radiation emitted to infinity.

\subsection{Radiation emitted to infinity}
\label{subsec_energy_to_infinity}
\begin{figure}
\center
\includegraphics[scale=0.45]{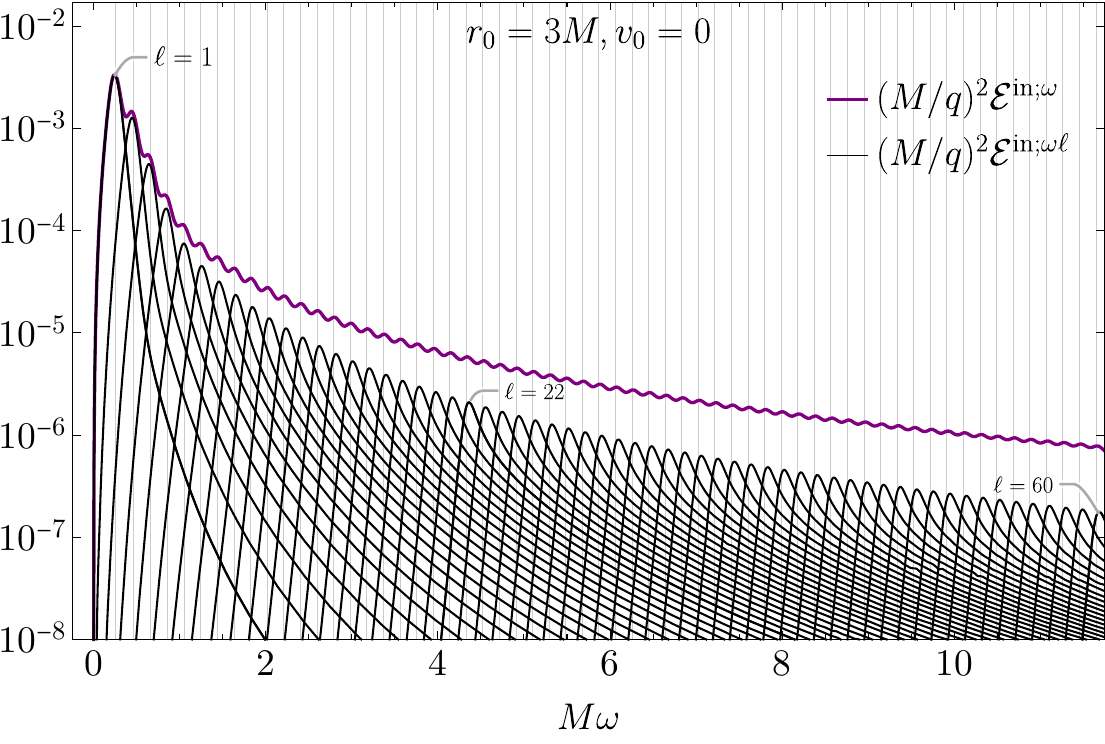}
\caption{The partial and total energy spectra, given by Eqs.~\eqref{energy_spectrum_sudd} and ~\eqref{total_energy_spectrum}, respectively, with $r_0 = 3M$ and $v_0=0,$ as a function of $M\omega.$ We consider the first $60$ multipoles.}
\label{to_infinity_r0_3M_v0_0}
\end{figure}

The spectrum of the radiation emitted to infinity is astrophysically relevant. It carries information about the BH and its vicinity. 
Figure~\ref{to_infinity_r0_3M_v0_0} shows the partial and total energy spectra, given by Eqs.~\eqref{energy_spectrum_sudd} and~\eqref{total_energy_spectrum}, with $r_0=3M$ and $v_0=0$, in a log plot.
 For these values of $r_0$ and $v_0$, we observe that the partial energy spectrum features a maximum approximately at the fundamental frequency of the BH quasinormal modes, $\omega_{\ell}^{\mathrm{qnf}}.$ The total energy released by the charged particle, as given by Eq.~\eqref{total_energy}, is $\mathcal{E}^{\mathrm{in}} \approx 0.0010 q^{2}/M.$ The partial emitted energy, given by Eq.~\eqref{partial_energy}, for the multipole number 
$\ell=1$ ($\ell=20$)
 corresponds approximately to
$55.25\%$ ($0.114\%$)
of $\mathcal{E}^{\mathrm{in}},$ the total energy emitted to infinity. Thus, the majority of the emitted energy comes from the dipole contribution. (The contribution of $\ell=1$ decreases for smaller $r_0$ and increases for larger $r_0,$ up to
about $83.23\%$ for $r_0 \to \infty.$) The dimensionless quantity $(M/q^{2})\mathcal{E}^{\mathrm{in}}$ corresponds approximately to $0.175\%$ of the specific energy $E$ of the charged particle given by Eq.~\eqref{energy}.

\begin{figure*}
\center
\includegraphics[scale=0.6]{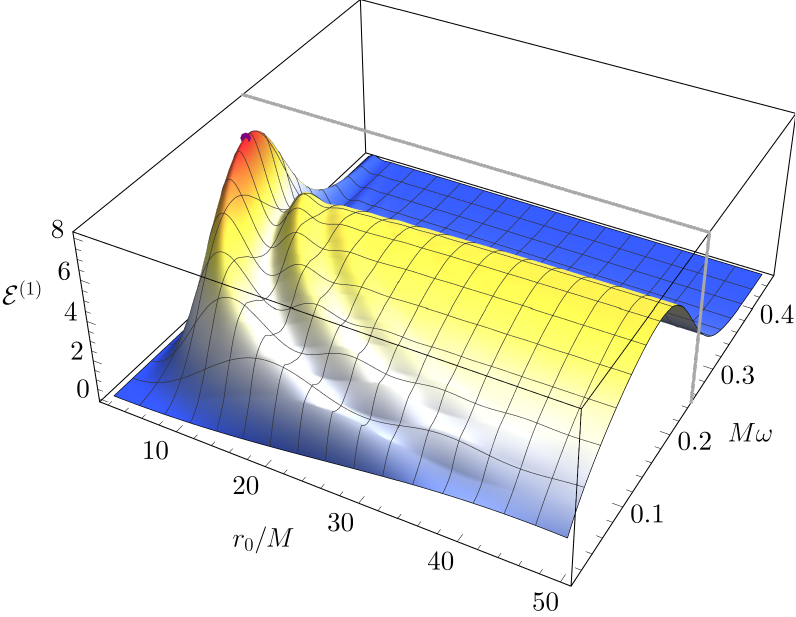}
\includegraphics[scale=0.6]{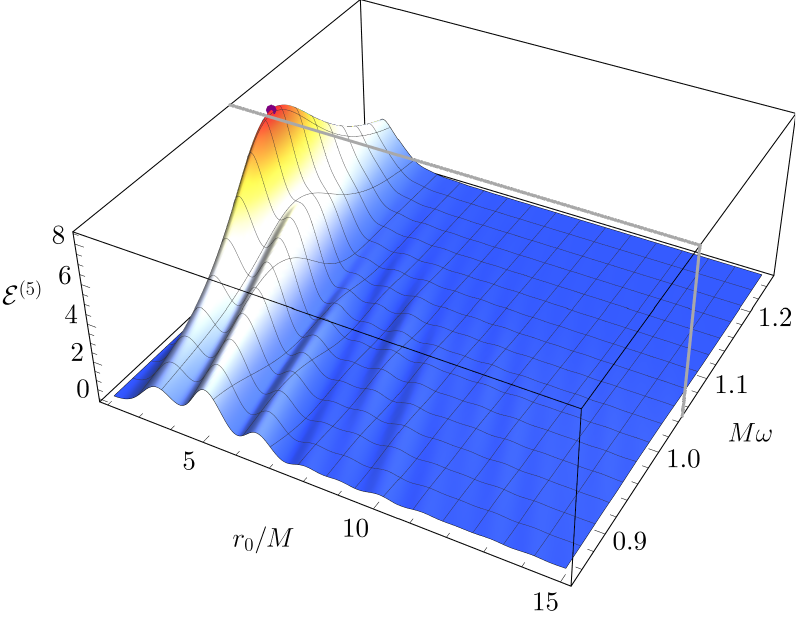}
\caption{The partial energy spectrum $\mathcal{E}^{(\ell)} \equiv 10^{\frac{3\ell}{\abs{\ell-2}}}(M/q)^{2}\mathcal{E}^{\mathrm{in};\omega \ell}$ depicted as a function of $r_0/M$ and $M\omega$, with $v_0=0,$ for $\ell=1$ (left) and $\ell=5$ (right). The (gray) facegrids mark the positions of $M\omega_{\ell}^{\mathrm{qnf}}.$ The (purple) disks mark the points of the global maximum of each function in the given range, at $r_0/M \approx 5.4504, 2.7598$ and $M\omega  \approx 0.1924, 1.0619,$ respectively.}
\label{fig_energy_spectrum_v0_0_l}
\end{figure*}

The behavior of the partial energy spectrum $\mathcal{E}^{\mathrm{in}; \omega  \ell}$ for $r_0$ finite and $v_0=0,$ shown in Fig.~\ref{to_infinity_r0_3M_v0_0}, is just for a representative value $r_0=3M$. For general values of $r_0,$ the maximum of the spectrum shifts around the quasinormal frequency. In general, as $r_0$ increases from around $3M,$ the peak shifts to $\omega<\omega_\ell^{\mathrm{qnf}}$; while as $r_0$ decreases, the peak shifts to $\omega>\omega_\ell^{\mathrm{qnf}}.$ However, for intermediate values of $r_0,$ i.e., for $4M \lesssim r_0 \lesssim 20M,$ the behavior is more complicated because the partial energy spectrum 
has multiple local maxima and minima (see Fig.~\ref{fig_energy_spectrum_v0_0_l}).

Figure~\ref{fig_energy_spectrum_v0_0_l} illustrates the partial energy spectrum for two representative values of $\ell$ ($\ell=1,5$), depicted as a function of $r_0/M$ and $M \omega$ (around $M \omega_{\ell}^{\mathrm{qnf}}$). We observe that as the multipole number $\ell$ increases, the corresponding value of $r_0$ associated with the maximum (partial) energy emission decreases, while the value of $\omega$ at the peak of the (partial) energy spectrum increases. The global maxima of the energy spectra for $\ell=2,3,4$ are located at $r_0/M \approx 3.6062, 3.0647, 2.8487$ and $M\omega \approx 0.4203, 0.6402, 0.8531,$ respectively. (These cases are not plotted in Fig.~\ref{fig_energy_spectrum_v0_0_l}.)

\begin{figure}
\center
\includegraphics[scale=0.45]{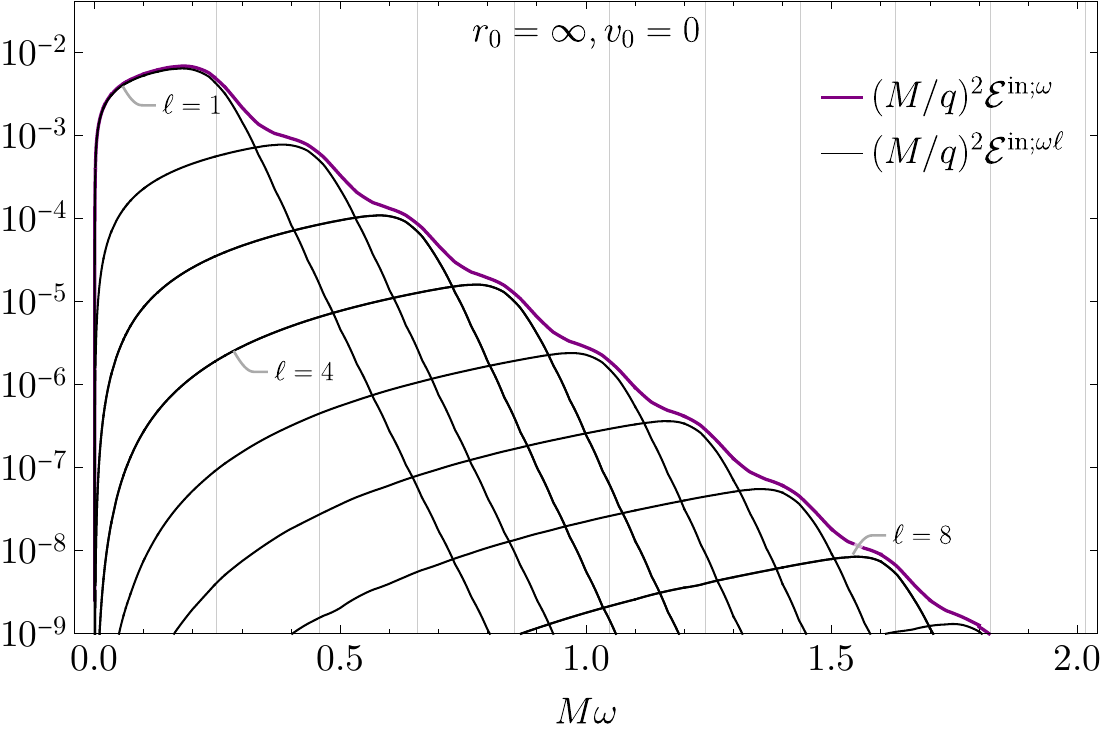}
\caption{The partial and total energy spectra, given by Eqs.~\eqref{energy_spectrum_sudd} and ~\eqref{total_energy_spectrum}, respectively, with $r_0 = \infty$ and $v_0=0,$ as a function of $M \omega.$ We consider the first $9$ multipoles.}
\label{to_infinity_r0_inf_v0_0}
\end{figure}

Next we discuss the emitted radiation for $r_0 \to \infty.$ Figure~\ref{to_infinity_r0_inf_v0_0} shows the partial and total energy spectra for the charged particle released from rest at infinity, i.e., for $v_0=0$ and $r_0 \to \infty.$ We see that the spectrum falls rapidly with increasing $\ell,$ and the total energy spectrum shows an exponential decay for $\omega>\omega_{\ell=1}^{\mathrm{qnf}}$, where $\omega_{\ell=1}^{\mathrm{qnf}} \approx 0.2482M.$ For $M\omega \to 0,$ we have 
$\mathcal{E}^{\mathrm{in};\omega \ell} \to 0$, confirming
an observation in Sec.~\ref{Inmoder0infty}. The total energy released to infinity from the charged particle, as given by Eq.~\eqref{total_energy}, is $\mathcal{E}^{\mathrm{in}} \approx 0.0017 q^{2}/M.$ The partial emitted energy, given by Eq.~\eqref{partial_energy}, for $\ell=2$ ($\ell=5$), corresponds approximately to $14.04\%$ 
($0.058\%$) of $\mathcal{E}^{\mathrm{in}}.$ The dimensionless quantity $(M/q^{2})\mathcal{E}^{\mathrm{in}}$ corresponds
approximately to $0.170 \%$ of the specific energy $E$ of the charged particle given by Eq.~\eqref{energy}.  Examining Fig.~\ref{to_infinity_r0_inf_v0_0} in comparison to Fig.~\ref{to_infinity_r0_3M_v0_0}, we observe a clear suppression of higher multipoles in the case $r_0 \to \infty$.

\begin{figure}
\center
\includegraphics[scale=0.45]{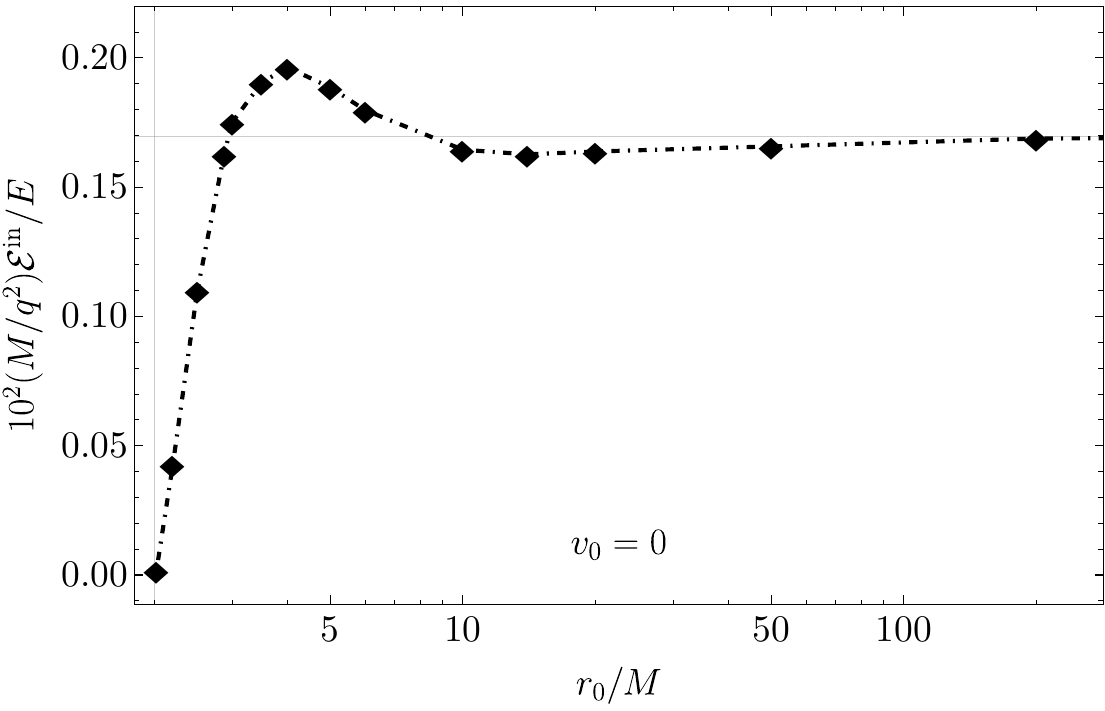}
\caption{The log-linear-scaled plot of the percentage of initial energy of the particle radiated to infinity, depicted as a function of $r_0/M,$ with $v_0=0.$ The vertical (gray) line marks the BH horizon, and the horizontal (gray) line marks the value of the quantity in the limit $r_0 \to \infty.$}
\label{Ratio_energy_to_infinity_to_initial_energy}
\end{figure}

The percentage of the initial energy released to infinity depends on $r_0,$ as shown in Fig.~\ref{Ratio_energy_to_infinity_to_initial_energy}. We see that  the percentage of the initial energy released has a global maximum at $r_0 \approx 4M$.
If the particle is released near the horizon, only a small portion of the initial energy is radiated to infinity.

\begin{figure}[h]
\center
\includegraphics[scale=0.45]{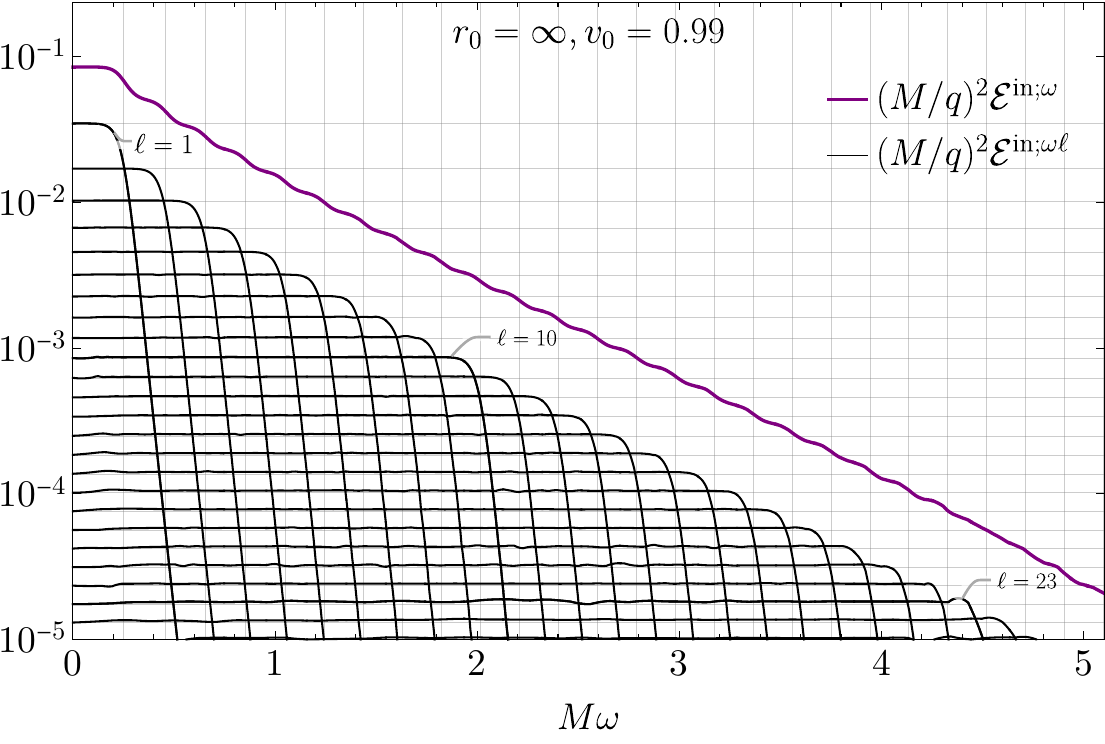}
\caption{The partial and total energy spectra, given by Eqs.~\eqref{energy_spectrum_sudd} and ~\eqref{total_energy_spectrum}, respectively, with $r_0 = \infty$ and $v_0=0.99,$ as a function of $M \omega.$}
\label{to_infinity_r0_inf_ultrarelat}
\end{figure}

Next, we discuss the influence of the nonzero initial velocity $v_0$ on the emitted energy spectrum for charges projected from infinity. Figure~\ref{to_infinity_r0_inf_ultrarelat} shows the partial and total energy spectra associated with an ultrarelativistic charged particle 
($v_0=0.99$ and $r_0 \to \infty$). By comparing it with Fig.~\ref{to_infinity_r0_inf_v0_0}, we see that the total spectrum decays more slowly for large $\omega$ for an ultrarelativistic particle. The $\omega \to 0$ limit of the partial
energy spectrum $\mathcal{E}^{\mathrm{in};\omega \ell}$
shown in Fig.~\ref{to_infinity_r0_inf_ultrarelat} agrees well with the analytic result, Eq.~\eqref{low_freq_partial_energy}, for each $\ell$, as we saw before. The partial emitted energy can be approximated using Eq.~\eqref{partial_energy_v0_approx_1}, which deviates from the numerical results by less than $3\%$.
The total energy released to infinity, as given by Eq.~\eqref{total_energy}, for the first $30$ multipoles, is $\mathcal{E}^{\mathrm{in}} \approx 0.0546 q^{2}/M$. This value is also in good agreement with the one obtained via Eq.~\eqref{partial_energy_v0_approx_1}, $\mathcal{E}^{\mathrm{in}} \approx 0.0540 q^{2}/M$.
We note, in particular, that the partial energy emitted for $\ell=1$ ($\ell=10$) represents approximately $16.25\%$ ($3.198\%$) of $\mathcal{E}^{\mathrm{in}}.$ 
That is, the dipole contribution is not dominant in this case. The dimensionless quantity $(M/q^2)\mathcal{E}^{\mathrm{in}}$ corresponds
approximately to $0.771 \%$ of the specific initial energy $E$ of the charged particle given by Eq.~\eqref{energy}.

\begin{figure}
\center
\includegraphics[scale=0.46]{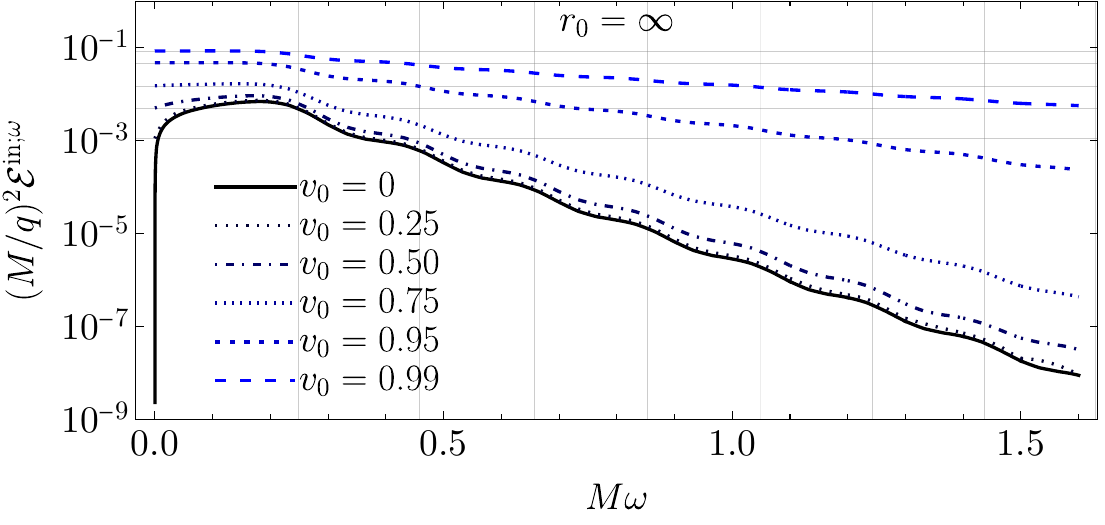}
\caption{The total energy spectrum, given by Eq.~\eqref{total_energy_spectrum}, for $r_0\to\infty,$ with some choices of $v_0,$ as indicated.}
\label{to_infinity_total_spectrum}
\end{figure}

As we observed before, from Eq.~\eqref{low_freq_partial_energy}, we see that the zero-frequency limit of $\mathcal{E}^{\mathrm{in};\omega \ell}$ is nonzero, provided that $v_0 \neq 0.$ 
As the initial velocity $v_0$ increases, the total energy spectrum decreases less and less rapidly for large $\omega$.  Figure~\ref{to_infinity_total_spectrum} illustrates the total energy spectrum $\mathcal{E}^{\mathrm{in}; \omega}$, for some choices of $v_0$ ranging from $0$ to values close to $1$. We see that the total emitted spectrum increases with $v_0.$ This is due to the excitation of higher multipoles, as Fig.~\ref{to_infinity_r0_inf_ultrarelat} indicates. 

In the next subsection, we analyze the electromagnetic energy absorbed by the BH.
\subsection{Electromagnetic energy absorbed by the black hole}
\label{to_horizon}
\begin{figure*}
\center
\includegraphics[scale=0.45]{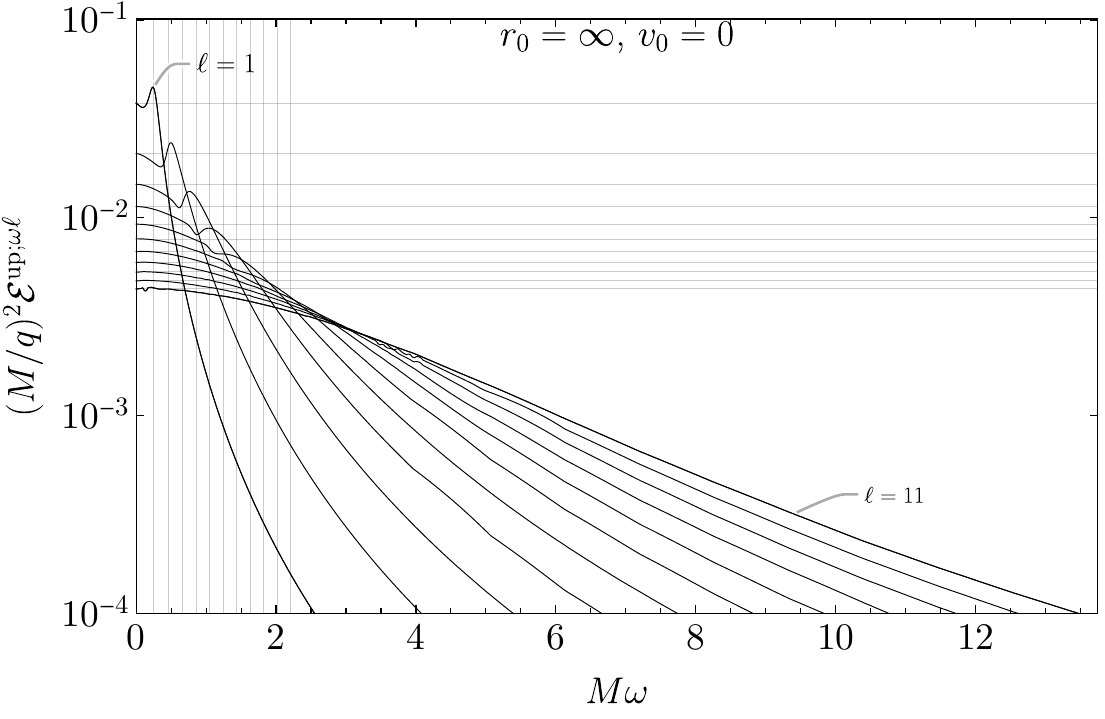}
\includegraphics[scale=0.45]{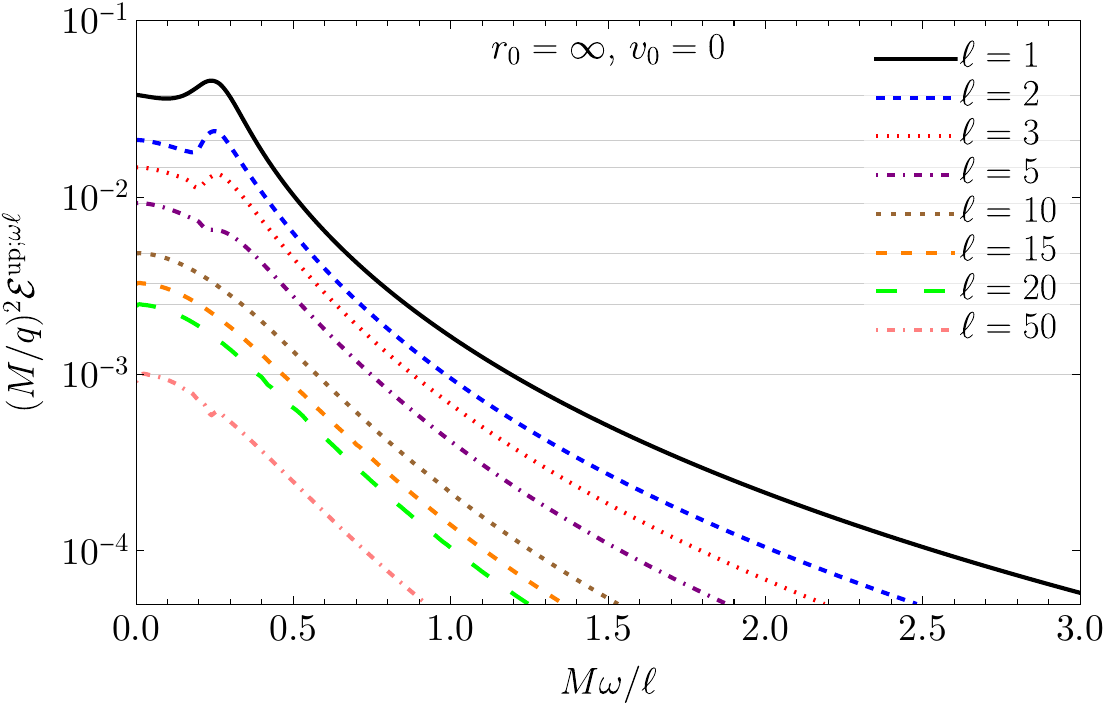}
\caption{The partial energy spectrum $\mathcal{E}^{\mathrm{up}; \omega \ell}$ is shown as a function of $M \omega$ (left) and $M \omega/\ell$ (right), with $r_0 = \infty$ and $v_0=0.$}
\label{to_horizon_r0_inf_v0_0}
\end{figure*}

Figure~\ref{to_horizon_r0_inf_v0_0} shows the partial energy spectrum absorbed by the BH for the first $10$ multipoles, plotted as a function of $M\omega,$ and also some higher multipoles, plotted as a function of $M\omega/\ell.$ The charge is released from rest at infinity. The contribution from each multipole $\ell$ is roughly constant: $\mathcal{E}^{\mathrm{up}; \ell} \approx 0.02 q^2/M,$ for higher multipoles. We also observe this behavior for finite values of $r_0,$ but the numerical value decreases like $\sqrt{f(r_0)}$ as 
$r_0$ decreases. Hence, the constant partial contribution is approximately written as $\mathcal{E}^{\mathrm{up}; \ell} \approx \sqrt{f(r_0)}\times 0.02 q^2/M$ for higher multipoles. For $r_0=4M$ we have $\mathcal{E}^{\mathrm{up}; \ell} \approx 0.014 q^2/M.$

The approximate $\ell$ independence of the partial energy
$\mathcal{E}^{\mathrm{up}; \ell}$
for large $\ell$ implies that the total electromagnetic
energy absorbed by the BH is infinite.
This infinity
suggests that the partial energies
$\mathcal{E}^{\mathrm{up}; \ell}$ for large $\ell$ represent 
the infinite Coulomb energy around the point charge, which
flows across the horizon when the charge falls into the
BH. As shown in the Appendix, this hypothesis
leads to the estimate $\mathcal{E}^{\mathrm{up};\ell} \approx Eq^2/16\pi M$ for large $\ell$, where the specific energy $E$ of the point charge is given by Eq.~\eqref{energy}. This formula
agrees with
our numerical results
quite well. (Note that $E=\sqrt{f(r_0)}$ if $v_0=0$.) We note that the Coulomb energy should be regarded as part
of the mass energy of the point charge by a ``classical
renormalization''. Thus, the partial energies
$\mathcal{E}^{\mathrm{up}; \ell}$ do not represent the
true radiation for large $\ell$.

\begin{figure}
\center
\includegraphics[scale=0.45]{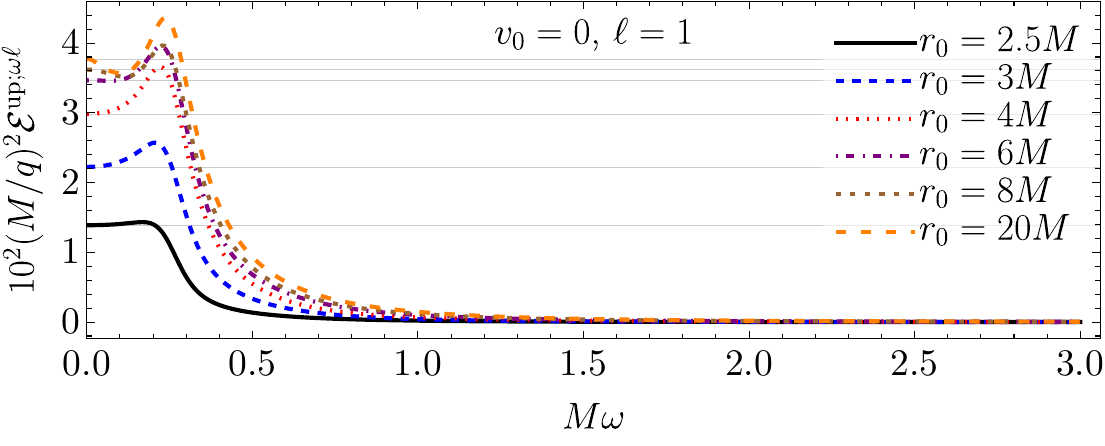}
\includegraphics[scale=0.45]{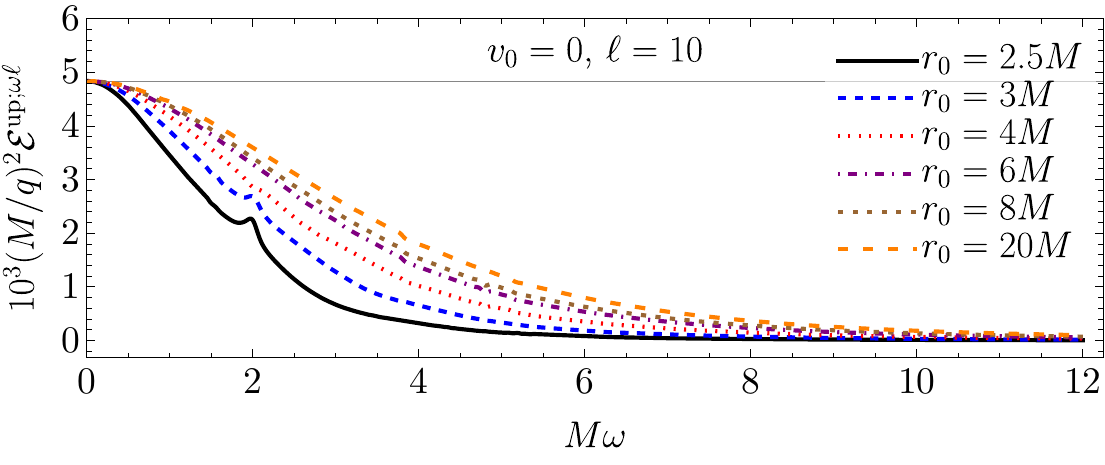}
\caption{The partial energy spectrum $\mathcal{E}^{\mathrm{up}; \omega \ell}$ for $\ell=1$ (top) and $\ell=10$ (bottom), with finite $r_0$ and $v_0=0.$}
\label{to_horizon_r0_finite_v0_0_l}
\end{figure}

Figure~\ref{to_horizon_r0_finite_v0_0_l} shows the partial energy spectrum with $v_0=0$, for two representative values of $\ell$ and different choices of $r_0.$ As $\ell$ increases, the spectrum at the zero-frequency limit converges to a value almost independent of $r_0$, as we
saw in Sec.~\ref{up-finite-r0}.

\begin{figure}
\center
\includegraphics[scale=0.45]{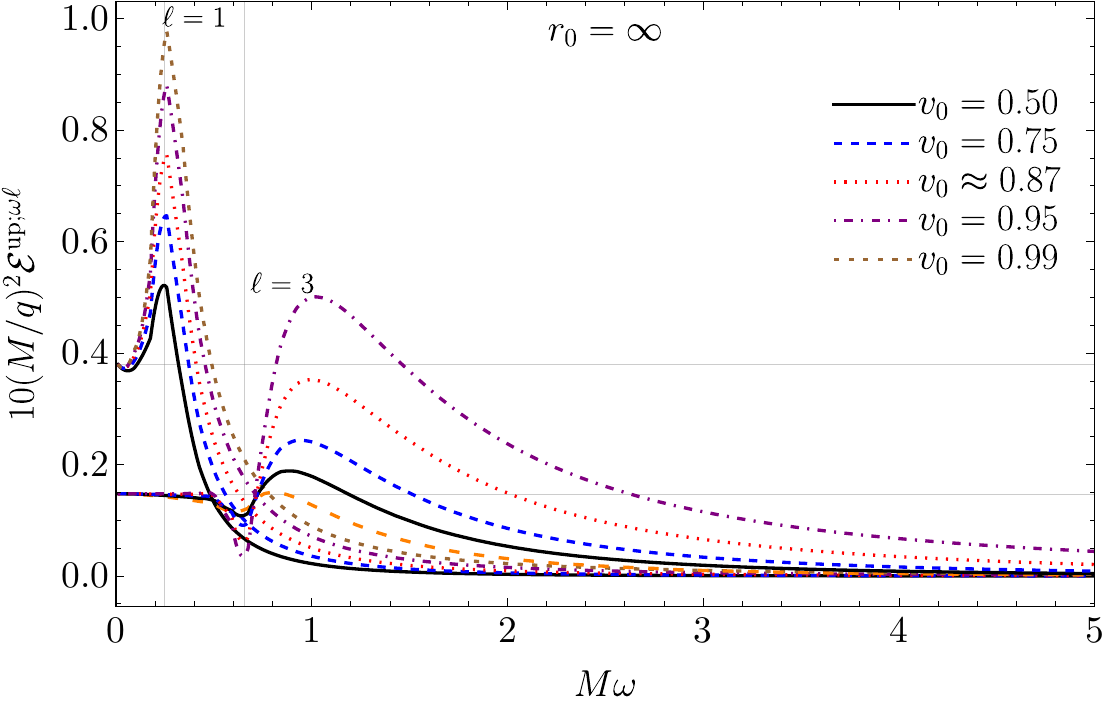}
\caption{The partial energy spectrum $\mathcal{E}^{\mathrm{up}; \omega \ell}$ with $\ell=1$ and $\ell=3,$ as a function of $M \omega,$ for $r_0 = \infty$ and some choices of $v_0 > 0.$}
\label{to_horizon_r0_inf_v0_not_0}
\end{figure}

Figure~\ref{to_horizon_r0_inf_v0_not_0} shows the partial energy spectrum for two choices of $\ell.$ The charge is projected from infinity with $v_0 > 0.$ The partial spectra
are found to be independent of $v_0$ in the low-frequency regime, as indicated by the overlapping curves in the left panel of Fig.~\ref{to_horizon_r0_inf_v0_not_0}.
This agrees with an observation made in Sec.~\ref{up-finite-r0}, which is also valid for $r_0=\infty$.

\subsection{Radiation due to an extended charge}
\label{extended}
In the previous subsection, the emitted energy going
into the horizon was analyzed
using the point particle approximation, which is an
idealization.
We showed that this idealization leads to the result that infinite electromagnetic energy
is absorbed by the BH. We show in the Appendix that this
infinite energy can be explained as the energy due to the
Coulomb field around the point charge, which should be
regarded as part of the mass energy of the point charge rather than the true radiation.

\begin{figure}
\center
\includegraphics[scale=0.45]{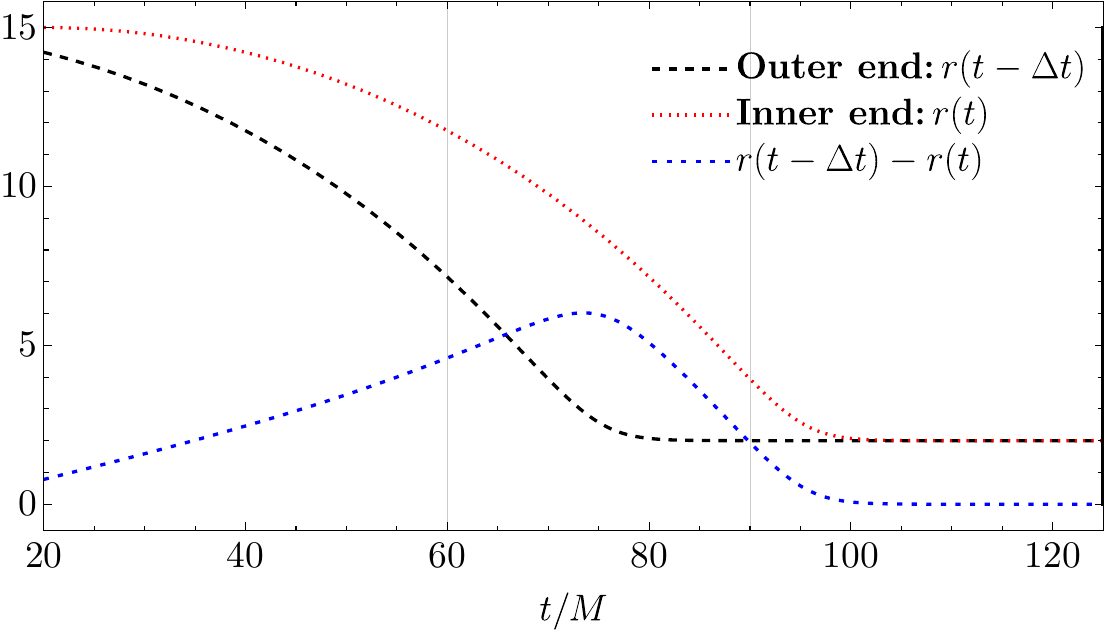}
\includegraphics[scale=0.22]{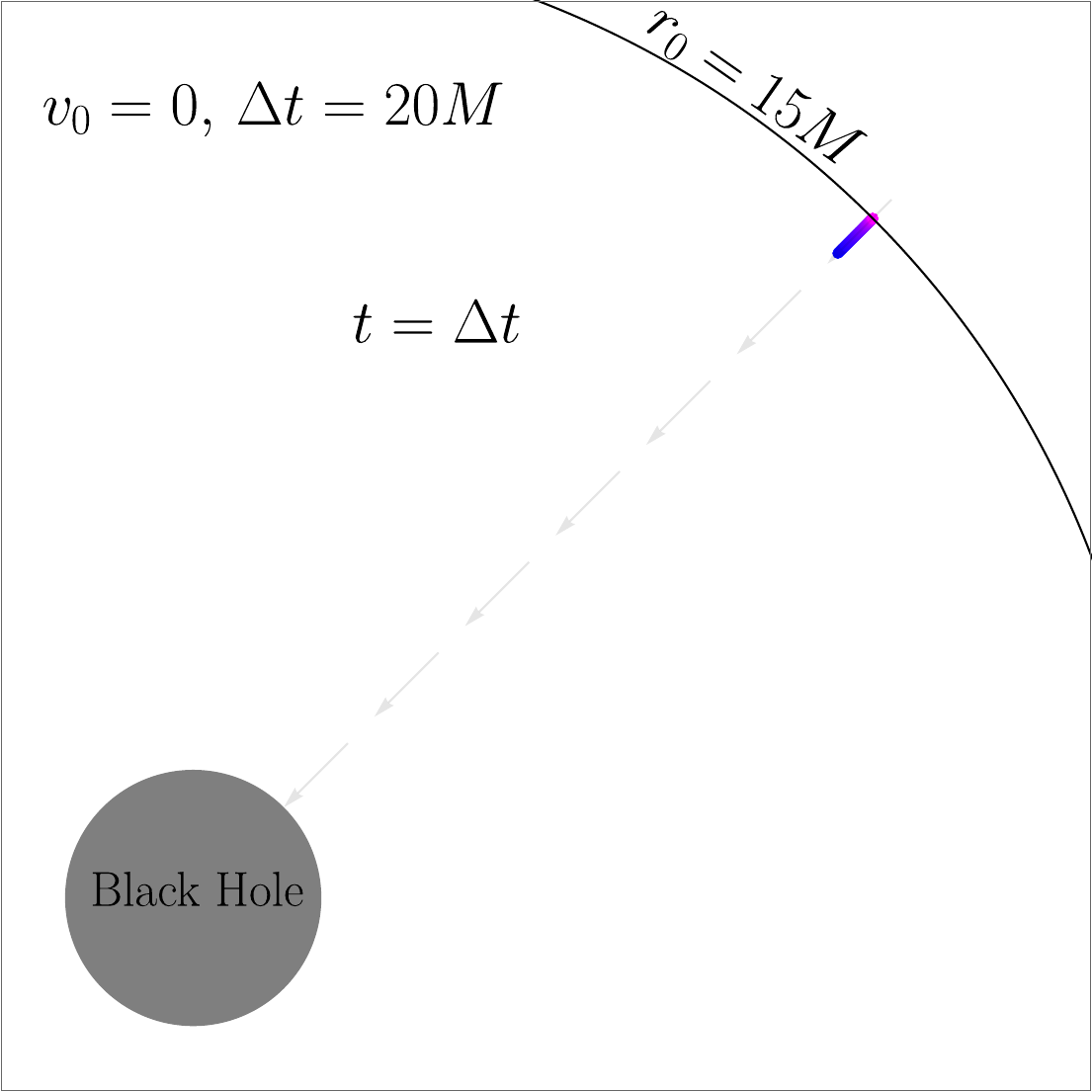}
\includegraphics[scale=0.22]{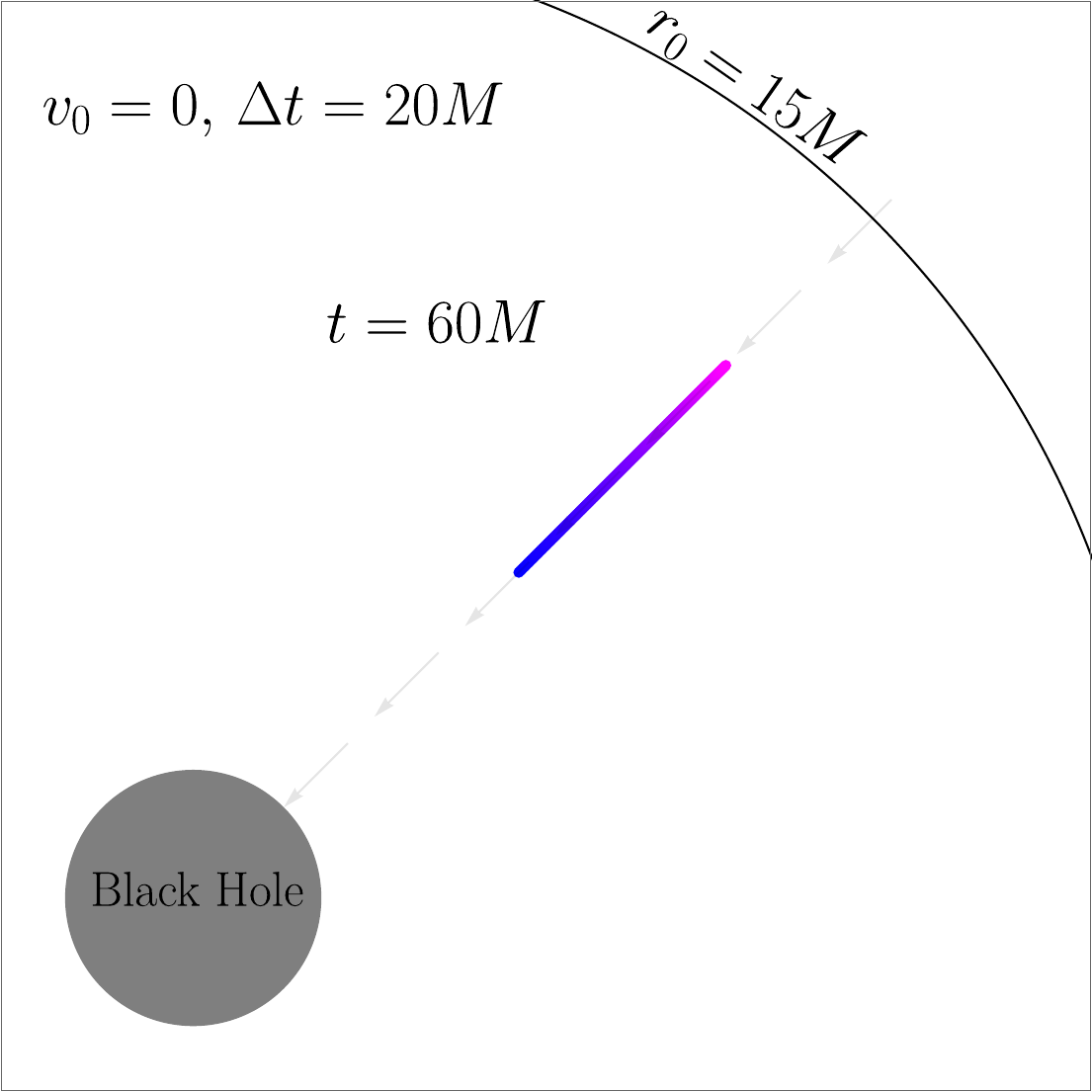}
\includegraphics[scale=0.22]{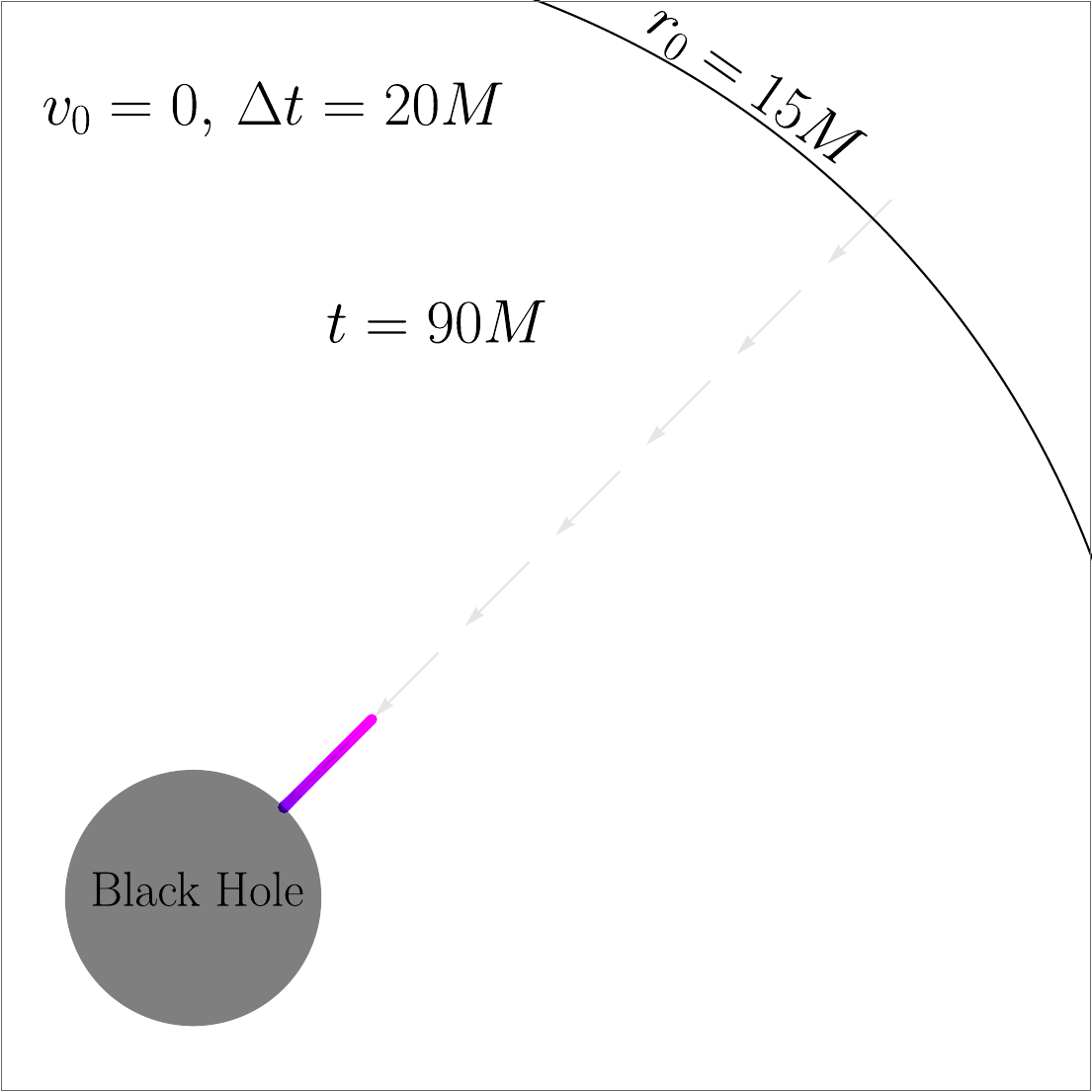}
\includegraphics[scale=0.22]{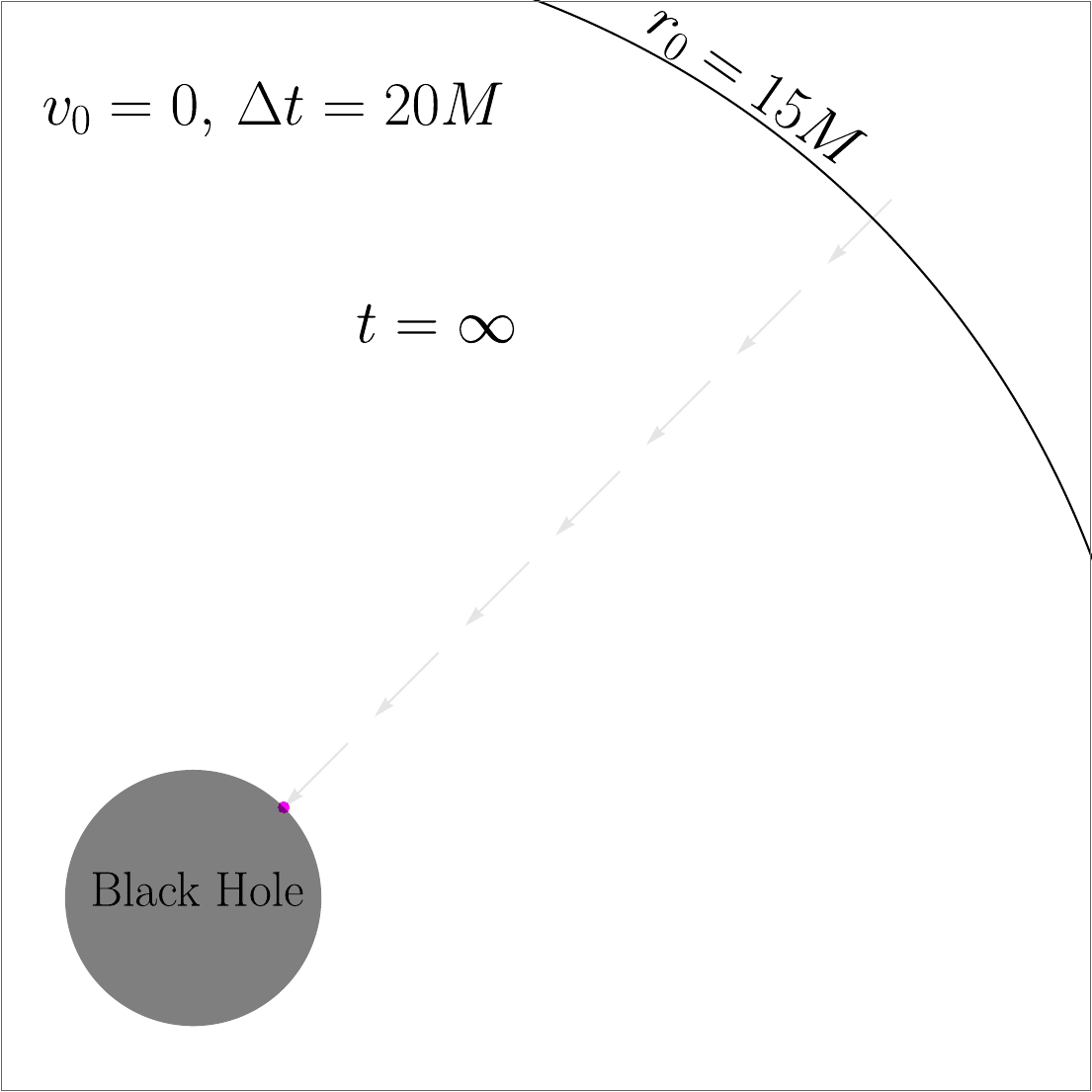}
\caption{Top: the radial positions of the inner and outer ends of the string as a function of the time $t.$ The difference $r(t-\Delta t)- r(t)$ is also plotted. Bottom: the process of the string falling radially into the BH is represented for four different values of the parameter $t,$ with $r_0=15M$ and $v_0=0.$}
\label{String_falling_graph}
\end{figure}

The infinity of the Coulomb energy is 
milder for an 
extended charged body than for a point charge. Therefore,
we expect that the partial energy 
$\mathcal{E}^{\mathrm{up};\ell}$ absorbed by the BH will decrease as a function of $\ell$ for an extended charged
body.
In this subsection, we study
a charged one-dimensional object extended in the radial direction (see, e.g., Refs.~\cite{tiomno_1972,ruffini_1972,nakamura_1981,haugan_1982,barausse_2021} and the references therein).
An extended object that is easy to implement 
numerically is a system of $N$ noninteracting particles, each with charge $q/N,$ distributed along the radial direction, such that all the charges (each labeled by $j$) follow the same radial geodesic [the one characterized by the specific energy $E$ given by Eq.~\eqref{energy}] and that the charge labeled by $j$ is
released from the same point, but later in time by the amount $j \Delta t/(N-1),$ where $\Delta t$ is a constant. In our semiclassical approach, this time shifting in the trajectory results in a different phase factor in the transition amplitude associated with each charge. This gives rise to interference between the radiation amplitudes due to individual charges.\footnote{See Ref.~\cite{mendes_2011} for an interesting investigation of radiation interference from scalar sources in circular orbits.}
One can readily see that the transition amplitude $\mathcal{A}_{N}^{n;\omega \ell m}$
associated with the $N$-charge system is given by
\begin{equation}
\label{amplitude_N}
\mathcal{A}_{N}^{n;\omega \ell m} = \left(\sum_{j=0}^{N-1} \frac{e^{i \omega \frac{j \Delta t}{N-1}}}{N}\right)\mathcal{A}^{n;\omega \ell m}.
\end{equation}
Note that when the charge with label $j=N-1$ is released, the charge with label $j=0$ (released at $t=0$)
is already located at $r(\Delta t)$ with inward velocity given by Eq.~\eqref{reciprocal_radial_velocity}.
This process is illustrated in the limit $N\to \infty$ in Fig.~\ref{String_falling_graph}.

\begin{figure}
\center
\includegraphics[scale=0.45]{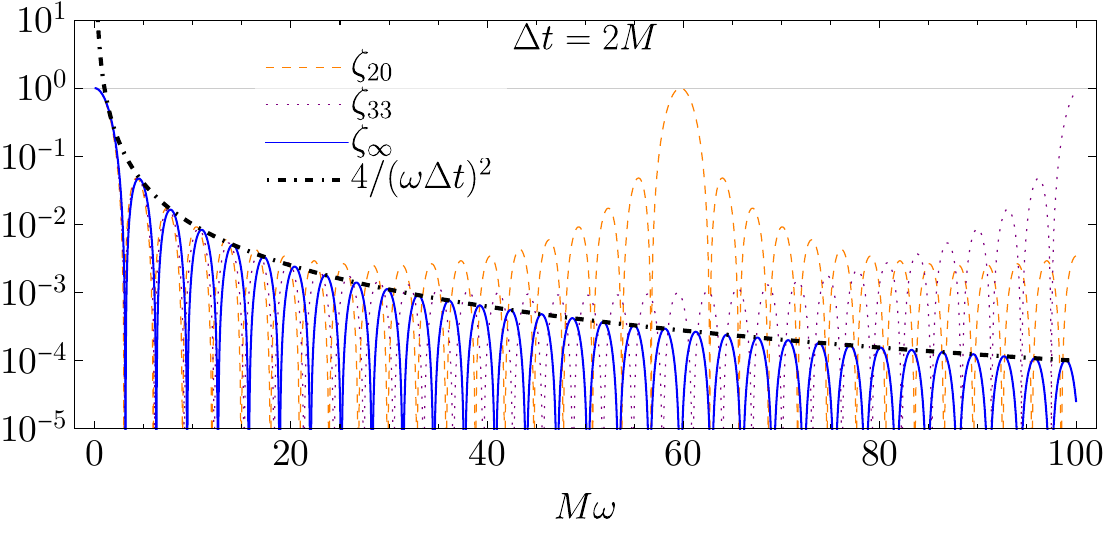}
\caption{The function $\upzeta_N(\omega),$ given in Eq.~\eqref{zeta-factor-discrete}, for some choices of $N$ and $\Delta t = 2M.$}
\label{zeta}
\end{figure}

The partial energy spectrum associated with the $N$-charge system can be written as
\begin{equation}
\label{partial_spectrum_N}
\mathcal{E}_{N}^{n; \omega \ell} = \upzeta_{N}(\omega)\mathcal{E}^{n; \omega \ell},
\end{equation}
where
\begin{equation}
\upzeta_{N}(\omega) =\left|\sum_{j=0}^{N-1} \frac{e^{i \omega \frac{j \Delta t}{N-1}}}{N}\right|^2. 
\label{zeta-factor-discrete}
\end{equation}
This is an oscillatory factor ranging between $0$ and $1$ and satisfying $\upzeta_N\boldsymbol{(}\omega+2\pi(N-1)/\Delta t\boldsymbol{)} = \upzeta_N(\omega)$.
Note that the point particle limit is obtained by letting $\Delta t \to 0$.
If we let $N \to \infty,$ the sum in
Eq.~\eqref{zeta-factor-discrete} becomes an integral.
Thus, we find
\begin{equation}
\label{zeta_inf}
\upzeta_{\infty}(\omega) = \left( \frac{2\sin \frac{\omega\Delta t}{2}}{\omega \Delta t} \right)^2.
\end{equation}
The factor $\upzeta_{\infty}(\omega)$ completely governs the overall behavior of the energy spectra in the high-frequency region. It decays as $4/(\omega\Delta t)^2$
for large $\omega$, as illustrated in Fig.~\ref{zeta}. 
The factor $\upzeta_N(\omega)$ is
a periodic function of $\omega$ for any finite $N$, and only in the
limit $N\to \infty$ does it decrease as a function of $\omega$
for all values of $\omega$.   Figure~\ref{to_horizon_r0_inf_v0_0} shows that the energy spectrum $\mathcal{E}^{\mathrm{up};\omega \ell}$ for the point charge
extends to higher and higher frequencies as the multipole number
$\ell$ increases.  Thus, to have $\mathcal{E}^{\mathrm{up};\ell}$ tend to $0$, instead of a nonzero constant, as $\ell\to\infty$ in the multicharge model it is crucial to take the limit $N \to \infty$, because
only then the high-frequency contribution to $\mathcal{E}^{\mathrm{up};\ell}$ is suppressed. Note that there will be no radiation with the frequencies at the zeros of 
$\upzeta_{\infty}(\omega).$ This observation applies to
the radiation emitted to infinity, as well as the
electromagnetic energy absorbed by the BH.
Thus,
there will be no radiation with the wavelengths $\Delta t/n$, $n=1,2,3,\ldots$.
For instance, one could eliminate from the emitted spectrum the $\ell$th fundamental quasinormal frequency of the BH by setting $\Delta t = 2 \pi n/\omega_\ell^{\mathrm{qnf}}.$

\begin{figure}
\center
\includegraphics[scale=0.45]{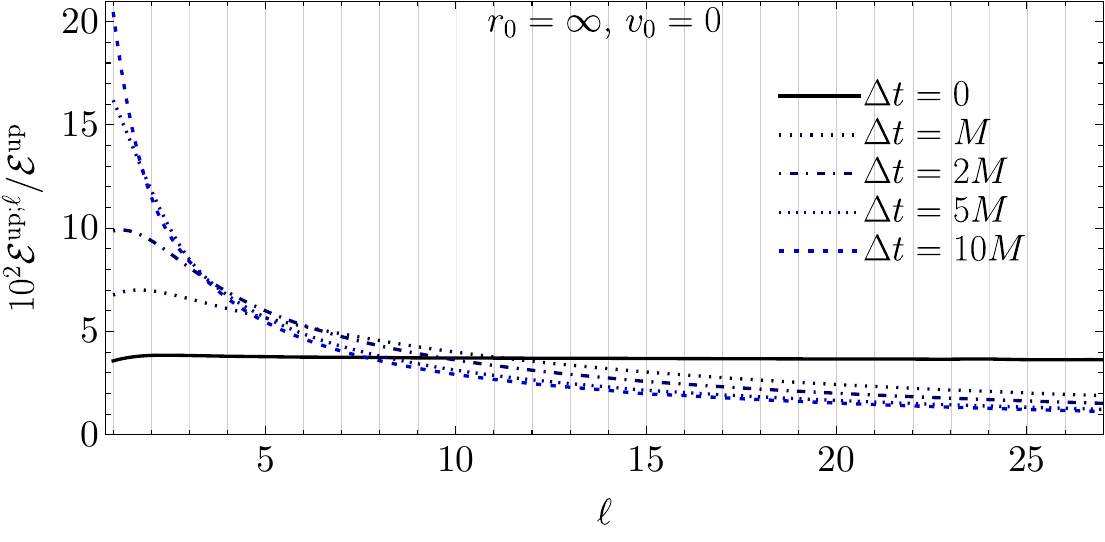}
\caption{The relative multipole contribution to the absorbed energy for a charged string released from $r_0=\infty,$ as the percentage in the sum up to $\ell=27$, denoted here by $\mathcal{E}^{\mathrm{up}}$,
as a function of the multipole number $\ell.$ The vertical gridlines indicate the values of $\ell.$}
\label{extend_up_multipole_contribution}
\end{figure}

\begin{figure}
\center
\includegraphics[scale=0.45]{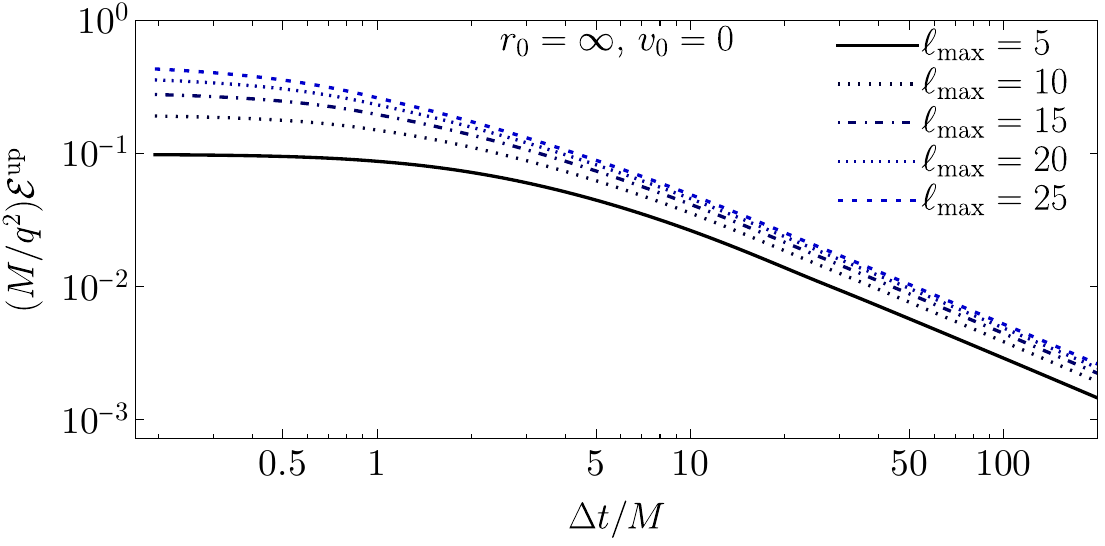}
\caption{The absorbed energy for a charged string released from $r_0=\infty,$ as a function of $\Delta t.$ The $\ell$ sum in Eq.~\eqref{total_energy} is truncated at $\ell_{\mathrm{max}}.$ Different choices of $\ell_{\mathrm{max}}$ are considered.}
\label{extend_up_multipole_contribution_tot}
\end{figure}

As was stated earlier, the contribution of each multipole to the total energy absorbed by the BH is roughly constant for large $\ell$ in the point particle model. This is no longer true for the string model, though the total energy absorbed remains infinite because the charge is still concentrated on a line.
When we increase the length $\Delta t$ from zero (point particle model), the relative contribution
of the lower multipoles increases, whereas that of the higher multipoles decreases. This can be seen in Fig.~\ref{extend_up_multipole_contribution}, where the partial energy is plotted against the multipole number $\ell$ for different choices of $\Delta t$.
This figure shows the contribution of each $\ell$ to $\mathcal{E}^{\mathrm{up}}$, defined as the sum of the absorbed partial
energies $\mathcal{E}^{\mathrm{up};\ell}$ up to $\ell=27$,
i.e., $\mathcal{E}^{\mathrm{up};\ell}/\mathcal{E}^{\mathrm{up}}$, where $\mathcal{E}^{\mathrm{up}} = \sum_{\ell=1}^{27}\mathcal{E}^{\mathrm{up};\ell}$, in percentage. 
In Fig.~\ref{extend_up_multipole_contribution_tot}, we show the quantity 
$\mathcal{E}^{\mathrm{up}} = \sum_{\ell=1}^{\ell_{\mathrm{max}}}\mathcal{E}^{\mathrm{up};\ell}$ as a function of $\Delta t$ for some
values of $\ell_{\mathrm{max}}$. Since the higher-multipole
contribution becomes smaller relatively to the lower-multipole contribution as $\Delta t$ increases, the
values of $\mathcal{E}^{\mathrm{up}}$ as defined above
for different $\ell_{\mathrm{max}}$ converge as $\Delta t$ increases,
as seen in this figure.

\begin{figure}
\center
\includegraphics[scale=0.45]{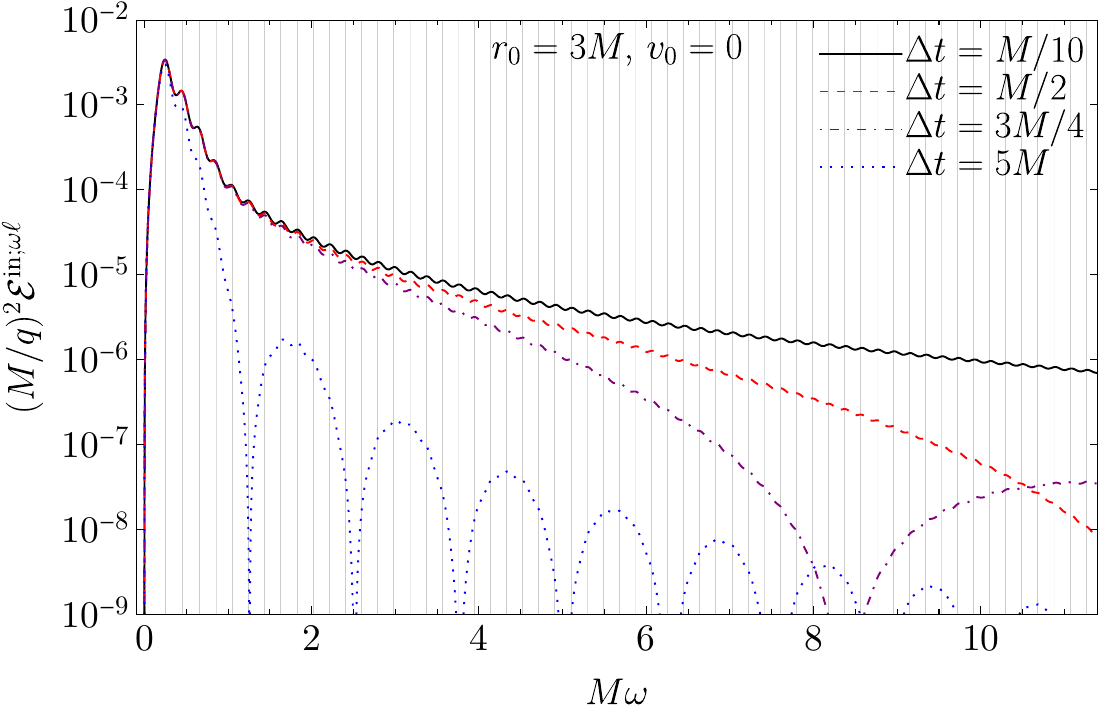}
\caption{The total energy spectrum for a charged string projected toward the BH from $r_0=3M$ with $v_0=0$ and some choices of $\Delta t$, as indicated.}
\label{spectrum_extended_in}
\end{figure}

The total energy spectrum for representative values of $r_0$ and $\Delta t$, for the radiation emitted to 
infinity, is shown in Fig.~\ref{spectrum_extended_in}. This figure can be compared with Fig.~\ref{to_infinity_r0_3M_v0_0}. We see that the spectrum is governed by $\upzeta_{\infty}(\omega)$ at high frequencies. In the limit $\Delta t \to \infty,$ we have $\upzeta_{\infty}(\omega) \to 0,$ and therefore, no radiation is emitted.

\section{Final Remarks}
\label{Sec_remarks}
In this paper, we analyzed the radiation emitted by a charge projected radially toward a Schwarzschild BH using quantum field theory in curved spacetime at tree level.
We confirm the results in Refs.~\cite{ruffini_1972,tiomno_1972,cardoso_eletr_2003,folacci_2020}, which use classical field theory, with additional insights and results. In particular, we obtained analytical results in the zero-frequency limit and were able to use a result in this limit to
find an approximate formula for the energy emitted to infinity by ultrarelativistic charges falling into the BH. We also presented a detailed analysis of the energy emitted by charges released from rest at a finite distance from the BH. We also provided a possible explanation for the recurring issue related to the divergence observed in the energy absorbed by the BH.

We verified that the radiation emitted to infinity is mostly of dipole origin for a charge falling from rest at $r_0 \gtrsim 3M,$ with increasing dipole contribution for increasing $r_0.$ The energy radiated is only a tiny fraction of the initial specific energy of the charge.
When the charge is projected with some initial velocity, higher multipoles are excited, and the total spectra become flatter. In this case, the emitted energy corresponds to a larger fraction of the specific initial energy of the charge. In particular, we confirmed for the ultrarelativistic case that the partial spectra
are approximately the flat spectra with a cutoff at the associated quasinormal frequency.
Thus, we showed that the analytic zero-frequency limit of the spectrum we derived in Sec.~\ref{Inmoder0infty} multiplied by the quasinormal frequency of the Schwarzschild BH 
gives a good approximation to each partial energy, and hence the total energy, emitted to infinity by the falling charge in ultrarelativistic motion.

We also studied radiation from a radially extended
charged ``string'' projected toward the BH.  This ``string'' is formed by $N$ noninteracting pointlike charges following the same radial geodesic, but they are released at different points in time, sequentially one after another. In the framework of quantum field theory, the sequential release of these charges introduces different phase factors for the probability emission amplitudes associated with each charge, thus producing interference.
As a result, in the $N \to \infty$ limit, we have an additional multiplier factor decaying like $\omega^{-2}$ in the energy spectrum. This factor is present in the energy spectrum emitted to infinity and to the horizon. This factor cuts off 
high-frequency contribution to the energy spectrum, thus
reducing the total energy radiated to infinity. This analysis is analogous to the classical analysis addressed in, e.g., Refs.~\cite{nakamura_1981,haugan_1982,barausse_2021}, for the case of gravitational radiation and yields similar results.

We also confirmed that the electromagnetic energy absorbed by the BH, for a pointlike charge released from rest, has approximately the same contribution from each multipole number $\ell$ for large $\ell$ and that the total absorbed energy diverges~\cite{tiomno_1972}, analogously to the gravitational case~\cite{davis_et_al_1972}. We found that this is also true for a charge released from a finite distance from the BH. Such divergence is a consequence of the point particle approximation. We showed that this
divergence can be explained as coming from the infinite Coulomb energy
around the point charge. (There is a similar divergence in the gravitational energy absorbed by the 
BH for a point mass falling
into a Schwarzschild BH~\cite{davis_et_al_1972}. This divergence can be shown to have 
the same explanation.) The same explanation must also hold for any particle trajectory that reaches the horizon. We also showed that the divergence is 
milder if the point charge is replaced by a one-dimensional extended charged body.

\begin{acknowledgments}
The authors thank Funda\c{c}\~ao Amaz\^onia de Amparo a Estudos e Pesquisas (FAPESPA),  Conselho Nacional de Desenvolvimento Cient\'ifico e Tecnol\'ogico (CNPq) and Coordena\c{c}\~ao de Aperfei\c{c}oamento de Pessoal de N\'{\i}vel Superior (Capes) - Finance Code 001, in Brazil, for partial financial support.
J. P. B. B. and L. C. B. C. thank the University of York, in England, and University of Aveiro, in Portugal, respectively, for the kind hospitality during the completion of this work.
This work has further been supported by the European Union's Horizon 2020 research and innovation (RISE) programme H2020-MSCA-RISE-2017 Grant No. FunFiCO-777740 and by the European Horizon Europe staff exchange (SE) programme HORIZON-MSCA-2021-SE-01 Grant No. NewFunFiCO-101086251.
\end{acknowledgments}

\appendix*
\section{\uppercase{Multipole decomposition of the Coulomb energy going into the horizon}} \label{app-Coulomb}

The near constancy of the contribution from each multipole number $\ell$ to the electromagnetic energy absorbed by the BH  for large $\ell$ implies that
the total electromagnetic energy absorbed by the BH is infinite.  We show in this appendix that this behavior of the energy absorbed
found numerically can be explained by the energy of the Coulomb field surrounding the charge about to fall into the BH.
The energy of the electric field around a point charge is infinite, and this infinity is due to the contribution from the
electric field arbitrarily close to the charge. Thus, we only need to analyze the electric field near the charge on the BH horizon. 

We first write down the Coulomb potential close to a charge $q$ on the horizon in the in-going Eddington--Finkelstein coordinate system, in which
the metric is given by
\begin{align}
    d\tau^2 & = \left(1-\frac{2M}{r}\right)dv^2 - 2drdv - r^2(d\theta^2 + \sin^2\theta\,d\phi^2),
    \label{EF-metric}
\end{align}
where $v$ is constant on each in-going radial null geodesic. The coordinate $v$ is related to the time coordinate $t$ in Schwarzschild 
coordinates by
\begin{align}
v = t + r + 2M\ln\frac{r-2M}{2M}.    
\end{align}

The vector $\partial_v$ is a Killing vector, and the corresponding
conserved quantity, the specific energy of a point particle falling in the radial direction, is given by
\begin{align}
    E & = \left( 1-\frac{2M}{r}\right) \frac{dv}{d\tau} - \frac{dr}{d\tau},
\end{align}
where $\tau$ is the proper time of the charge.
The specific energy $E$ in terms of the initial position and velocity is given by Eq.~\eqref{energy}.
The world line of the charge satisfies
\begin{align}
    \frac{dv}{d\tau} & = \frac{1}{E + \sqrt{E^2 - 1+2M/r}},\\
    \frac{dr}{d\tau} & = - \sqrt{E^2 - 1+2M/r}.
\end{align}
At the instant when the charge is on the horizon $r=2M,$ we have
$(dv/d\tau,dr/d\tau) = (1/2E, -E)$.  The radial line perpendicular to the world line of the charge with respect to the metric~\eqref{EF-metric} 
satisfies $(dv/ds,dr/ds) = (1/2E,E)$, where $s$ is the proper distance.

Now we define the local time and space coordinates, $\eta$ and $\rho$, near the charge on the horizon such that 
$(d\eta/d\tau,d\rho/d\tau) = (1,0)$ on the world line of the charge, and
$(d\eta/ds,d\rho/ds)=(0,1)$ on the radial line perpendicular to this world line.  Choosing $\eta=\rho=v=0$ at the horizon on the world line of the charge, we find approximately
\begin{align}
    \eta & = E v - \frac{1}{2E}(r-2M),\\
    \rho & = E v + \frac{1}{2E}(r-2M).
\end{align}
We note that the Killing vector $\mathcal{K}^\mu = (\partial_v)^\mu$ has components $\mathcal{K}^\eta = \mathcal{K}^\rho = E$ in the local $\eta$-$\rho$ coordinate system
(with $\mathcal{K}^\theta=\mathcal{K}^\phi = 0$) at the position of the charge
on the horizon.

The vector potential $A_\mu$ near the charge on the horizon ($r=2M$) is approximately the $\eta$-independent Coulomb field,
\begin{align}
    A_\eta = \frac{q}{4\pi R},\label{Coulomb-potential}
\end{align}
where
\begin{align}
    R & = \sqrt{\rho^2 + (2M)^2 \theta^2}, \label{R-approx1}
\end{align}
with all other components vanishing.
(We are using the equality sign rather imprecisely here.)  Near the charge, i.e., for $\rho \ll 2M$ and $\theta \ll 1$, we may replace this expression with
\begin{align}
    R & = \sqrt{a^2 - 2ab\cos\theta + b^2}, \label{R-approx2}
\end{align}
where
\begin{align}
    a & = \frac{1}{2}(\sqrt{\rho^2 + 4(2M)^2} + |\rho|),\\
    b & = \frac{1}{2}(\sqrt{\rho^2 + 4(2M)^2} - |\rho|).
\end{align}
The expression for $R$ in Eq.~\eqref{R-approx2} vanishes only for $(\rho,\theta)=(0,0)$ and reduces to that given in Eq.~\eqref{R-approx1} for $\theta \ll 1$.
Therefore, we may use Eq.~\eqref{R-approx2} in Eq.~\eqref{Coulomb-potential} for estimating the contribution to the Coulomb energy from large $\ell$, 
since the infinite energy arises exclusively from the electric field near the charge.  

Then, with the definition
\begin{align}
    |\rho| = 4M\sinh s,
\end{align}
we find, using the standard generating function for the Legendre polynomials~\cite[Eq.~8.921]{gradshteyn},
\begin{align}
    A_\eta & = \frac{q}{4\pi a}\sum_{\ell=0}^\infty \left(\frac{b}{a}\right)^{\ell} P_{\ell}(\cos\theta)\notag \\
    & = \frac{q}{8\pi M}\sum_{\ell=0}^\infty e^{-(2\ell+1)s} P_{\ell}(\cos\theta),
\end{align}
where $P_{\ell}(x)$ is the Legendre polynomial of order $\ell$.
The nonvanishing components of the field-strength tensor $F_{\mu\nu} = \nabla_\mu A_\nu - \nabla_\nu A_\mu$ are 
\begin{align}
    F_{\eta\rho} & = \pm \frac{q}{16\pi M^2}\sum_{\ell=0}^\infty (\ell+1/2)e^{-(2\ell+1)s} P_{\ell}(\cos\theta),\\
    F_{\eta\theta} & = - \frac{q}{8\pi M}\sum_{\ell=0}^\infty e^{-(2\ell+1)s} \frac{d\ }{d\theta} P_{\ell}(\cos\theta),
\end{align}
where the plus sign is for $\rho > 0,$ and the minus sign is 
for $\rho < 0$.  We have let 
$\partial/\partial |\rho| \approx (4M)^{-1}\partial/\partial s$, since
we only need to estimate the singular behavior of the electromagnetic energy density near $\rho =0$ ($s=0$).

The stress-energy tensor is
\begin{align}
    T_{\mu\nu} & = - F_{\mu\alpha}{F_\nu}^\alpha + \frac{1}{4}g_{\mu\nu}F^{\alpha\beta}F_{\alpha\beta}.
\end{align}
The conserved energy-momentum current is $T_{\mu\nu}\mathcal{K}^\nu$.  The $\eta$-component of this covector is
\begin{align}
  T_{\eta\nu}\mathcal{K}^\nu 
    & = \frac{E q^2}{2(16\pi M^2)^2}\left\{ \left[\sum_{\ell=0}^\infty (\ell+1/2)e^{-(2\ell+1)s} P_{\ell}(\cos\theta)  \right]^2\right. \notag \\
    &\quad \quad 
    \left. + \left[\sum_{\ell=0}^\infty e^{-(2\ell+1)s}\frac{d\ }{d\theta}P_{\ell}(\cos\theta)\right]^2\right\}. \label{before-integral}
\end{align}
We integrate this quantity over the hypersurface $\Sigma_\eta$ of constant $\eta$, with the volume element 
\begin{align}
    d\rho\, r^2\sin\theta d\theta d\phi
    & \approx 16M^3\sin\theta dsd\theta d\phi,
\end{align}
where we have made the approximations $r^2\approx 4M^2$ and
$\cosh s \approx 1$.
The integral over $\rho$ near $\rho=0$ is replaced by 
twice the integral over $s$ from $0$ to $\infty$. (Again, 
we are interested
only in the contribution from small $|\rho|$ with large $\ell,$ and hence, the upper limit of the $s$-integral is not important.)
Then, we find, using the standard orthogonality relations
satisfied by the Legendre polynomials,  
that the infinite Coulomb energy can formally be expanded
as
\begin{align}
    \mathcal{E}^{\textrm{Coulomb}} & = \int_{\Sigma_\eta} T_{\eta\nu}\mathcal{K}^\nu  d\rho\, r^2\sin\theta d\theta d\phi \notag \\
    & = \sum_{\ell=0}^\infty \mathcal{E}_{\ell}^{\textrm{Coulomb}},
\end{align}
where 
\begin{align}
    \mathcal{E}_{\ell}^{\textrm{Coulomb}} \approx \frac{E q^2}{16\pi M}\ \ \textrm{for}\ \ell \gg 1. 
    \label{app-final-result}
\end{align}


\end{document}